\documentclass[review]{elsarticle}

\usepackage[left=2.5cm, right=2.5cm, top=2cm]{geometry}
\usepackage{tensor}
\usepackage{empheq}
\usepackage[table,xcdraw]{xcolor}
\usepackage{booktabs}
\usepackage{subcaption}
\usepackage{amssymb}
\usepackage[ruled,vlined]{algorithm2e}
\usepackage{listings}
\usepackage{hyperref}
\hypersetup{
    colorlinks,
    linkcolor={blue!100!black},
    citecolor={red!85!black},
    urlcolor={blue!80!black} }
\usepackage{lineno}
\modulolinenumbers[5]
\usepackage{lipsum} 
\usepackage[makeroom]{cancel}
\usepackage{yfonts}
\usepackage{gensymb}
\usepackage[toc,page]{appendix}

\usepackage{tikz}
\usepackage{pgfplots}

\journal{arXiv}

\newcommand{\norm}[1]{\left\lVert#1\right\rVert}

\newcommand{\D}{\textbf{D}}
\newcommand{\E}{\textbf{E}}
\newcommand{\F}{\textbf{F}}
\newcommand{\I}{\textbf{I}}
\newcommand{\Hb}{\textbf{H}}
\newcommand{\Cb}{\textbf{C}}
\newcommand{\A}{\textbf{A}_1}
\newcommand{\Aii}{\textbf{A}_2}
\newcommand{\Aiii}{\textbf{A}_3}
\newcommand{\f}{\textbf{f}}
\newcommand{\LL}{{\textbf{L}_1}}
\newcommand{\Li}{{L_1}}
\newcommand{\Lii}{{L_2}}
\newcommand{\Liii}{{L_3}}
\newcommand{\x}{\textbf{x}}

\newcommand{\xb}{\textbf{x}}

\newcommand{\ub}{\textbf{u}}
\newcommand{\dx}{\text{d}\textbf{x}}
\newcommand{\PP}{\textbf{P}}

\newcommand{\G}{\textbf{G}}

\newcommand{\B}{\textbf{B}}
\newcommand{\V}{\textbf{V}}
\newcommand{\Vd}{\textbf{U}}
\newcommand{\etab}{\boldsymbol{\eta}}
\newcommand{\thb}{\boldsymbol{\theta}}
\newcommand{\xib}{\boldsymbol{\xi}}
\newcommand{\Xib}{\boldsymbol{\Xi}}
\newcommand{\Phib}{\boldsymbol{\Phi}}
\newcommand{\fpp}[2]{\frac{\partial #1}{\partial #2}}

\newcommand{\ten}[3]{\tensor[_{#2}]{#1}{_{#3}}}
\newcommand{\tenb}[3]{\tensor[_{#2}]{\textbf{#1}}{_{#3}}}

\definecolor{lightgray}{gray}{0.75}

\DeclareRobustCommand{\hwplotA}{\raisebox{2pt}{\tikz{\draw[black, solid, line width=1.2pt](0,0) -- (5mm,0);}}}
\DeclareRobustCommand{\hwplotB}{\raisebox{2pt}{\tikz{\draw[red, dashed, line width=1.2pt](0,0) -- (5mm,0);}}}
\DeclareRobustCommand{\hwplotC}{\raisebox{2pt}{\tikz{\draw[green, solid, line width=1.2pt](0,0) -- (5mm,0);}}}

\DeclareRobustCommand{\hwplotD}{\raisebox{.5pt}{\tikz{\fill[blue!50] (0,0) circle (.075);\fill[blue!50] (.2,0) circle (.075);\fill[blue!50] (.4,0) circle (.075);}}}

\definecolor{hotmagenta}{rgb}{1.0, 0.11, 0.81}

\DeclareRobustCommand{\ROMd}{\raisebox{2pt}{\tikz{\draw[black, solid, line width=2pt](0,0) -- (5mm,0);}}}
\DeclareRobustCommand{\Nzero}{\raisebox{2pt}{\tikz{\draw[green, solid, line width=0.5pt](0,0) -- (5mm,0);}}}
\DeclareRobustCommand{\Nunot}{\raisebox{2pt}{\tikz{\draw[blue, solid, line width=0.5pt](0,0) -- (5mm,0);}}}
\DeclareRobustCommand{\Nuno}{\raisebox{2pt}{\tikz{\draw[red, dashed, line width=0.5pt](0,0) -- (5mm,0);}}}
\definecolor{darkgreen}{rgb}{0.07, 0.3, 0.1}
\definecolor{lightblue}{rgb}{0.30196, 0.7451, 0.93333}
\definecolor{darkred}{rgb}{0.63529, 0.078431, 0.18431}
\DeclareRobustCommand{\Nzerov}{\raisebox{2pt}{\tikz{\draw[darkgreen, dashed, line width=2pt](0,0) -- (5mm,0);}}}
\DeclareRobustCommand{\Nunotv}{\raisebox{2pt}{\tikz{\draw[lightblue, solid, line width=2pt](0,0) -- (5mm,0);}}}
\DeclareRobustCommand{\Nunov}{\raisebox{2pt}{\tikz{\draw[darkred, dashed, line width=2pt](0,0) -- (5mm,0);}}}

\usepackage{mathtools}

\begin{document}

\begin{frontmatter}

\title{An enhanced parametric nonlinear reduced order model for imperfect structures using Neumann expansion}

\author[poliaddress]{Jacopo Marconi}
\author[ethaddress]{Paolo Tiso\corref{mycorrespondingauthor}}
\cortext[mycorrespondingauthor]{Corresponding author}
\ead{ptiso@ethz.ch}
\author[poliaddress]{Davide E. Quadrelli}
\author[poliaddress]{Francesco Braghin}
\address[poliaddress]{Department of Mechanical Engineering, Politecnico di Milano, Via La Masa 1, 20156 Milan, Italy}
\address[ethaddress]{Institute for Mechanical Systems, ETH Z\"{u}rich, Leonhardstrasse 21, 8092 Z\"{u}rich, Switzerland}

\begin{abstract}
We present an enhanced version of the parametric nonlinear reduced order model for shape imperfections in structural dynamics we studied in a previous work \cite{Marconi2020}. The model is computed intrusively and with \textit{no training} using information about the nominal geometry of the structure and some user-defined displacement fields representing \textit{shape defects}, i.e. small deviations from the nominal geometry parametrized by their respective amplitudes. The linear superposition of these artificial displacements describe the defected geometry and can be embedded in the strain formulation in such a way that, in the end, nonlinear internal elastic forces can be expressed as a \textit{polynomial} function of both these defect fields and the actual displacement field. This way, a tensorial representation of the internal forces can be obtained and, owning the reduction in size of the model given by a Galerkin projection, high simulation speed-ups can be achieved. We show that by adopting a rigorous deformation framework we are able to achieve better accuracy as compared to the previous work. In particular, exploiting \textit{Neumann expansion} in the definition of the Green-Lagrange strain tensor, we show that our previous model is a lower order approximation with respect to the one we present now. Two numerical examples of a clamped beam and a MEMS gyroscope finally demonstrate the benefits of the method in terms of speed and increased accuracy.
\end{abstract}


\end{frontmatter}


\section{Introduction}
The Finite Element (FE) method has long been a fundamental analysis and design tool in many areas of science and engineering. In structural mechanics it is almost mandatory to use FE models to investigate the behavior of complex systems, which often have many geometric details that would be difficult to handle with alternative approaches, such as lumped parameter or analytical models \cite{Belytschko2014}. However, large FE simulations would often require considerable computational resources and time, so in some cases designers may prefer to perform real experiments rather than numerical ones. On the one hand, this need for fast and affordable FE simulations has given rise to numerical techniques to improve computational efficiency: domain decomposition and substructuring \cite{Toselli2005,Klerk2008} and FE Tearing and Interconnecting (FETI, \cite{Farhat1991}) are just a few examples. On the other hand, model order reduction methods have emerged, consisting in the construction of a Reduced Order Model (ROM), whose number of degrees of freedom (dofs) is much smaller than that of the reference Full Order Model (FOM). The use of linear ROMs also in industrial contexts is nowadays well established as the theory underlying them. Guyan reduction \cite{GUYAN1965} and \textit{modal analysis} \cite{He2001} are two well-known examples in mechanical statics and dynamics, respectively, where FOM's static deformations and Vibration Modes (VMs, also known as eigenmodes or natural modes of the linear system) are used to construct a Reduced Basis (RB) that projects the governing equations onto a lower dimension subspace. Linear ROMs were also successfully coupled with substructuring techniques in the Craig-Bampton and Rubin methods \cite{Craig1968, Rubin1975}, which are available in many commercial software. 

For nonlinear FE studies, where the demand for reduction is dire, many solutions have been proposed over the last decades, but none of them seems to have prevailed over the others, as each of them offers certain advantages, requires certain costs and/or targets specific problems. Overall, however, the literature is mature enough to provide the analyst with many different options in several practical applications, ranging from bolted joints \cite{Pichler2019}, gears \cite{Blockmans2015}, contacts \cite{Balajewicz2015, Geradin2016}, friction \cite{MehrdadPourkiaee2019} and viscoplasticity \cite{Ghavamian2017} to flexible multi-body dynamics with geometric nonlinearities \cite{Wu2019} and substructuring \cite{Wu2018}.

Nonlinear ROMs can be classified according to (i) whether they are RB-projection based or not, (ii) whether they are data- or model-driven and (iii) their (non-)intrusiveness. In the following we consider mostly projection approaches, as the one adopted in this work; alternatively, one could resort to different strategies, such as normal form theory or Spectral Submanifolds. The most recent contributions in this sense include \cite{Vizzaccaro2020} and \cite{Jain2018ssm, Ponsioen2020}. In (ii), for data-driven ROMs we usually refer to ROMs constructed using previous FOM simulation data (or experimental data, \cite{Perez2017}), as opposed to model-driven methods that rely on some intrinsic properties of the model itself for ROM construction, such as modal approaches \cite{Touze2014, Hollkamp2008, Kuether2015, Amabili2013}. As for intrusiveness, we usually denote a ROM as non-intrusive \cite{Mignolet2013} if it can be used with routines and solvers of commercial FE software and, conversely, as intrusive a method requiring dedicated routines. Specifically, intrusive methods require access and manipulation to element-level quantities, as for instance nonlinear generalized forces and jacobians. Other distinctions can be made in terms of the types of nonlinearities that a given model can handle and the way nonlinear functions are evaluated \cite{Jain2015}. All these differences ultimately affect the two phases that all ROMs have in common: the \textit{offline} phase, in which the ROM is constructed, and the \textit{online} phase, in which the simulation responses are retrieved. As the main goal of ROMs is to reduce computational effort and time, a key aspect to keep in mind when choosing a method is the overhead cost to pay in the offline phase; in the case of data-driven methods, this cost can be as high as the cost associated to the solution of the FOM \cite{Balajewicz2015}. Generally speaking then, data-driven methods (usually based on Proper Orthogonal Decomposition, or POD, strategies \cite{Lu2019}) are used in scenarios where the high cost associated to the data generation can be amortized: typically, this is the case of multi-query analysis. In this sense, although not as versatile and generally applicable as data-driven POD-based approaches, model-driven strategies in structural dynamics are desirable, for no FOM simulation is required a priori. Rayleigh-Ritz procedures \cite{Noor1980}, dual modes \cite{Mignolet2013} and Modal Derivatives (MDs) \cite{Idelsohn1985, Sombroek2018, Jain2017} are some popular examples.

One way to mitigate the offline overhead costs of all the aforementioned methods, but especially the data-driven ones, is to resort to (nonlinear) parametric ROMs, (NL-)pROMs. Also in this context, the literature on linear systems is quite well developed and consolidated. An extensive survey and comparison of these methods can be found in \cite{Benner2015, BaurBenner2017}. The reduction of nonlinear parametric Partial Differential Equations (PDEs) is instead still an active research topic, which has attracted increasing interest in various disciplines over the years. Interestingly, the vast majority of nonlinear parametric model order reduction methods is data-driven, POD-based. Some recent examples include non-intrusive interpolation methods for evaluating nonlinear functions with hypersurfaces \cite{Xiao2015, Xiao2017} and use of Gaussian Processes and machine learning for error evaluation and refinement of the pROM \cite{Xiao2019} or interpolation on the Grassman manifold via tangent spaces \cite{Zimmermann2019}. Alternatively, many of these methods approximate the nonlinear function using \textit{hyper-reduction} methods as the Discrete Empirical Interpolation Method (DEIM) \cite{Barrault2004, Chaturantabut2010} to speed up the evaluation, and in this sense online basis selection and adaptive algorithms were studied \cite{Phalippou2020, Cho2020}. However, as mentioned above, POD (and DEIM) needs a number of FOM simulations to construct the ROM. For this reason, \cite{Kast2020} implemented a Multi-Fidelity strategy in which the parametric dependence was reconstructed using a large number of low-fidelity models and a minimal number of high-fidelity evaluations. Other approaches exploit machine learning to construct an input-output relationship, with convolutional neural networks \cite{Hesthaven2018} and autoencoders \cite{Maulik2020}, which require the training of a network, again, using preexisting data. Note that most of the above methods lead to pROMs that are only \textit{evaluated} in the online phase, i.e. no simulation is actually performed\footnote{By \textit{simulation} we refer to the solution of a set of equations describing a system in any kind of analysis setting (e.g. in time or frequency domain).}, but the solutions at the known parameter locations are ``interpolated" to obtain the result.

Although model-driven NL-pROMs seem to be less popular, they offer the undeniable advantage of being simulation-free, thus considerably cutting down the offline costs. Interesting recent examples are loosely based on the extension of methods for linear systems, such as the Non-Linear Moment Matching (NLMM) scheme \cite{Astolfi2008, Astolfi2010, Ionescu2016}. In Ref. \cite{Rafiq2020}, a non-parametric ROM is constructed with NLMM and DEIM for each parameter instance sampled from the parameter space. These models are then ``adjusted" onto a common subspace where they are interpolated to produce the pROM. This strategy, however, requires the solution of a set of nonlinear algebraic equations on the FOM at different time instances, for different signal generators, and at each point on the parameter grid. For large systems, the computational effort could still be significant, although lower than that of POD methods.

In this paper we propose a NL-pROM for \textit{geometric nonlinearities} and parametrized \textit{shape defects} to study the behavior of imperfect structures. This is motivated by the fact that, as it is observed in many engineering applications, even small imperfections can significantly change characteristics and performances of a system, as for instance in the case of MEMS devices \cite{Acar2008, Izadi2018} and mistuning of gas turbine blades \cite{MehrdadPourkiaee2019}. Other ROMs have already been developed in this sense \cite{Wang2018a, Wang2018b}, but limited to \textit{localized} defects. Regarding geometric nonlinearities, we recall that in the case of continuum finite elements with linear elastic constitutive law and Total Lagrangian formulation, as in our study, the nonlinear elastic forces are a polynomials which can be represented using tensors, so that qualitatively\footnote{Due to memory limitations, third and fourth order stiffness tensors cannot be computed for the FOM, but they can be constructed in reduced form directly operating at element level \cite{Jain2015}.} the FOM governing equations write\footnote{$\otimes$ denotes the outer product, : and $\scalebox{.75}{ $\boldsymbol{\vdots}$ }$ the double and triple contraction operations.}
\begin{equation}
    \textbf{M}\ddot{\ub}^F + \textbf{C}_d\dot{\ub}^F + \tenb{K}{2}{}^F \ub^F + \tenb{K}{3}{}^F : (\ub^F \otimes \ub^F) + \tenb{K}{4}{}^F \scalebox{.75}{ $\boldsymbol{\vdots}$ } (\ub^F \otimes \ub^F \otimes \ub^F) = \f_{ext}(t)
\end{equation}
where \textbf{M}, $\textbf{C}_d \in \mathbb{R}^{n \times n}$ are the mass and damping matrices, $\ub^F, \,\dot \ub^F, \,\ddot \ub^F \in \mathbb{R}^n$ the displacement, velocity and acceleration vectors, and $\f_{ext}(t)\in \mathbb{R}^n$ an external forcing, being $n$ the FOM number of dofs. $\tenb{K}{2}{}^F \in \mathbb{R}^{n \times n}$, $\tenb{K}{3}{}^F \in \mathbb{R}^{n \times n \times n}$ and $\tenb{K}{4}{}^F \in \mathbb{R}^{n \times n \times n \times n}$ are the stiffness tensors for the linear, quadratic and cubic elastic internal forces.

\begin{figure}[t]
\centering
\includegraphics[width=.99\textwidth]{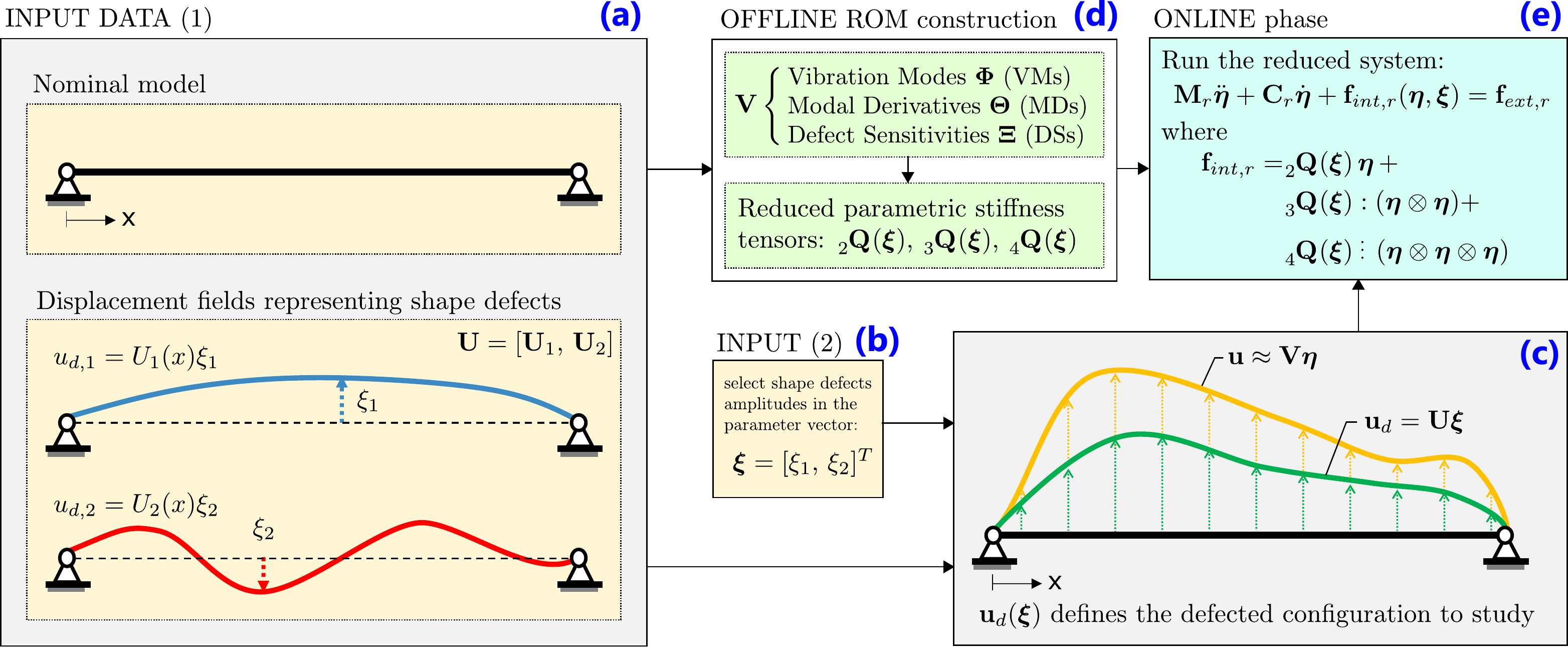}
\caption{Overview of the proposed method, schematically illustrated for a pinned beam.}
\label{fig:intro_scheme}
\end{figure}

Conceptually, the method retrace the one we presented in \cite{Marconi2020}, but it is based on a different deformation scheme (of which our earlier work resulted to be a sub-case). An overview of the individual steps of the method is shown in Fig. \ref{fig:intro_scheme}. The user defines as input data the nominal structure (in terms of geometry, material properties and FE mesh) and a number $m$ of displacement fields representing the \textit{shape defects}, which are intended as small deviations from the nominal geometry (Fig. \ref{fig:intro_scheme}a). These can be known analytically, from experimental measurements or previous simulations, and finally they can be discretized with displacement field vectors $\Vd_i$ and collected in a matrix $\Vd=[\Vd_1,...,\Vd_m]$. Each defect can then be leveraged in amplitude by the parameter vector $\xib=[\xi_1, ...,\xi_m]^T$ (Fig. \ref{fig:intro_scheme}b) so that the final defected geometry represented by the model is given by the global defect displacement field $\ub_d=\Vd\xib$, i.e. a linear superposition of the selected defects (Fig. \ref{fig:intro_scheme}c). With this information about the nominal structure and shape defects, we assemble the RB using a modal approach with VMs, MDs and Defect Sensitivities (DSs). We then construct the reduced stiffness tensors, \textit{once and for all}, projecting the element-level tensors with the selected RB (Fig. \ref{fig:intro_scheme}d). In this way, linear, quadratic and cubic elastic forces can be evaluated directly with respect to the reduced coordinates \emph{and} shape defect magnitudes without switching between the full and reduced order space when evaluating the nonlinear function. Our strategy can then be classified as model-driven (simulation-free). Finally, in the online phase, the simulation is performed with the reduced governing equations (Fig. \ref{fig:intro_scheme}e). Notice that the model is used to \textit{run} a simulation, not to \textit{evaluate} a solution as in interpolation-like techniques: as such, different forcing terms and also different analysis types (e.g. transient, frequency response) are possible.

All of this is possible thanks to the modified definition of the Green-Lagrange strain tensor we use. Specifically, our strain tensor embeds two subsequent transformations: (i) the one from nominal to defected geometry (which, at the end, will be parametrized), and (ii) the one from the defected configuration to the deformed/final one. The deformation produced by the latter is the one we measure, so no strain/stresses are introduced by the presence of the defect in (i); however, the deformation of (ii) will depend on (i). The formulation we obtain however contains rational terms which cannot be used for a tensorial representation (which can describe polynomials only). Given the assumed small entity of the shape defects, we advocate the use of a Neumann expansion to approximate the Green-Lagrange tensor, obtaining again a polynomial form. Applying standard FE procedures, we finally get to the expression of the reduced elastic internal forces, which will parametrically depend on the defect amplitudes $\xib$. In this framework, we show that the model in \cite{Marconi2020} (whose deformation formulation was based on \cite{Budiansky1967}) corresponds to a lower order Neumann expansion with integrals evaluated on the nominal volume, and that the higher order approximation we propose here leads to better accuracy and to a larger applicability range.

The work is organized as follows: the modified strain formulation is given in Section \ref{sec:2} and approximated using Neumann expansion in Section \ref{sec:3}. In Section \ref{sec:4} the FE discretization is developed and then used in Section \ref{sec:S5} to construct the reduced order stiffness tensors. The choice and computation of the RB is described in Section \ref{sec:6}. Finally, numerical studies in Sections \ref{sec:7} and \ref{sec:8} demonstrate the effectiveness of the proposed approach on a 2D FE clamped beam and on a MEMS gyroscope and computational times are discussed.

\section{Strain formulation: a two-steps deformation approach}
\label{sec:2}
%
\begin{figure}[t]
\centering
\includegraphics[width=.7\textwidth]{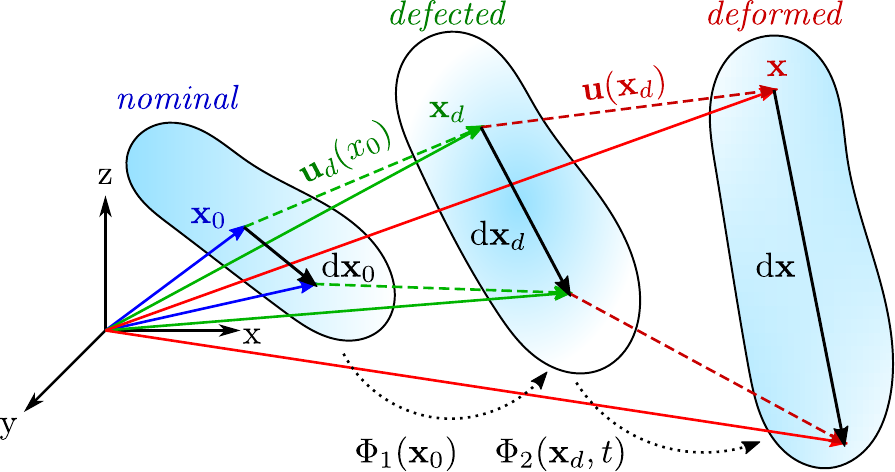}
\caption{Scheme for the considered deformation setting. A nominal structure, of coordinates $\x_0$, undergoes a deformation described by the transformation map $\Phi_1$. The structure is now in the deformed configuration (coordinates $\x_d$). A second transformation $\Phi_2$ and the displacement $\ub$ describe the deformation from the defected configuration to the final one.}
\label{fig:defscheme}
\end{figure}
Let us consider the scheme depicted in Fig. \ref{fig:defscheme}. A \textit{nominal} body of coordinates $\x_0=\{x_0,\,y_0,\,z_0\}$ undergoes a first deformation described by the map $\Phib_1$, which brings the body in a new configuration with coordinates $\x_d=\{x_d,\,y_d,\,z_d\}=\Phib_1(\x_0)$. The displacement corresponding to this operation is $\ub_d=\{u_d,\,v_d,\,w_d\}=\x_d-\x_0$. We will refer to this second configuration as the \textit{defected configuration}. As it will be detailed later, in our method $\ub_d$ will be a user-defined displacement field representing a shape defect which, superimposed to the nominal geometry, defines the configuration with respect to which we measure deformation. Let us now consider a second deformation, described by the map $\Phib_2$, from the defected configuration to the final one, with coordinates $\x(t)=\{x(t),\,y(t),\,z(t)\}=\Phib_2(\x_d,t)$. We will refer to the latter as to the \textit{deformed or final configuration}, whose displacement is given by $\ub=\{u,\,v,\,w\}=\x-\x_d$.

Considering the infinitesimal line segment $\dx_0$ in the nominal geometry, we can define the line segments d$\xb_d$ and d$\xb$ in the defected and deformed configurations as
\begin{subequations}
\begin{align}
    & \dx_d = \F_1\dx_0, \\
    & \dx = \F_2\dx_d=\F_2\F_1\dx_0,
\end{align}
\label{eq:fibers}
\end{subequations}
where the deformation gradients $\F_1$ and $\F_2$ are given by
\begin{subequations}
\begin{align}
    & \F_1 = \nabla_0 \xb_d = \fpp{\x_d}{\x_0} = \I + \nabla_0 \ub_d = \I + \fpp{\ub_d}{\x_0} = \I + \D_d, \\
    & \F_2 = \nabla_d \xb   = \fpp{\x  }{\x_d} = \I + \nabla_d \ub   = \I + \fpp{\ub}{\x_d} = \I + \D_2.
\end{align}
\label{eq:Fgradients}
\end{subequations}
and where $\D_d$ and $\D_2$ are the displacement derivative matrices of the first and second transformations, respectively. Using the chain rule, we can also define
\begin{equation}
    \D = \frac{\partial \ub}{\partial \x_0} = \frac{\partial \ub}{\partial \x_d} \frac{\partial \x_d}{\partial \x_0} = \D_2\F_1,
\label{eq:D2toD}
\end{equation}
so that $\D_2=\D\F_1^{-1}$ can be referred to the nominal coordinates.

Using Eqs. \eqref{eq:fibers}--\eqref{eq:D2toD}, the \textit{stretch} between deformed and defected configurations writes
\begin{equation}
\begin{split}
    \mathcal{S}=\dx^T\dx - \dx_d^T\dx_d &= \dx_0^T \F_1^T(\F_2^T\F_2 - \I)\F_1\dx_0 \\
    & = \dx_0^T (\D+\D^T+\D^T\D+\D_d^T\D+\D^T\D_d)\dx_0.
\end{split}
\label{eq:stretch}
\end{equation}
Measuring the deformation with respect to the \textit{defected} configuration, the second order Green-Lagrange strain tensor $\E_2$ is defined as
\begin{equation}
    \mathcal{S}=2\dx^T_d \E_2 \dx_d =2\dx^T_0 \F_1^T \E_2 \F_1 \dx_0,
\end{equation}
which, rearranged, leads to
\begin{equation}
    \E_2 = \frac 12 \F_1^{-T}(\D+\D^T+\D^T\D+\D_d^T\D+\D^T\D_d)\F_1^{-1}.
    \label{eq:E2}
\end{equation}
Looking at Eqs. \eqref{eq:stretch} and \eqref{eq:E2}, it can be easily verified that $\E_2$ correctly satisfies the minimum requirements for a strain measure to vanish under a rigid body translation ($\F_2=\I$) and/or rotation ($\F_2^T\F_2 = \textbf{R}^T\textbf{R}=\I$, with $\textbf{R}$ orthonormal rotation matrix), \textit{for any} $\F_1$. Eq. \eqref{eq:E2} is indeed an exact expression for the strains from defected to final configuration. Notice however that in this form all the quantities are computed with respect to the nominal coordinates $\xb_0$.

\section{Strain approximations}
\label{sec:3}
The introduced strain measure, being referred to the nominal geometry only, paves the way for the pre-computation of the stiffness tensors, as it will be shown in the following sections. However, as mentioned in the introduction, a tensorial formulation can be applied only when the internal forces display a polynomial dependence on the displacements, which in the present case include both $\ub_d$ and $\ub$. The inverse of the deformation gradient $\F_1$ in Eq. \eqref{eq:E2} entails a rational dependence on $\ub_d$, and therefore needs some attention. Let us consider the following known result:
\paragraph{Neumann expansion}
If $\PP$ is a square matrix and the Neumann series $\sum_{n=0}^{+\infty} \PP^n$ is convergent, we have that
\begin{equation}
    (\I-\PP)^{-1} = \sum_{n=0}^{+\infty} \PP^n
\end{equation}
A spectral norm\footnote{The spectral norm of a matrix $\A$ is defined as the square root of the largest singular value of $\A^*\A$, being $\A^*$ the conjugate transpose of $\A$, that is: $ \norm{\A}_2 = \sqrt{ \lambda_{max}(\A^*\A)} $} $\varepsilon=\norm{\PP}_2<1$ is a \textit{sufficient} condition for the convergence of the Neumann series. Moreover, it can be shown \cite{Wang2013} that truncating the sum to order $N$ the norm error is bounded as
\begin{equation}
\delta_N=\norm{\left(\I-\PP\right)^{-1}-\sum_{n=0}^{N} \PP^n}_2 \leq \dfrac{\varepsilon^{N+1}}{1-\varepsilon} = \delta_{lim}.
\label{eq:delta}
\end{equation}

Letting $\PP=-\D_d$, we can expand $\F_1^{-1}$ using the Neumann series as
\begin{equation}
\F_1^{-1} = (\I+\D_d)^{-1} \approx \sum_{n=0}^{N} (-\D_d)^n.
\label{eq:F1expansion}
\end{equation}

Under the assumption of small defects (i.e. $\norm{\D_d}_2 \ll1$) the series is guaranteed to converge. Moreover, we can truncate the sum in Eq. \eqref{eq:F1expansion} to $N=1$, obtaining:
\begin{equation}
\E_{2,N} = \frac 12 (\I - \D_d)^T(\D+\D^T+\D^T\D+\D_d^T\D+\D^T\D_d)(\I - \D_d),
\label{eq:E2N}
\end{equation}
which, solving the product, can be rewritten as:
\begin{multline}
\E_{2,N} = \frac 12 \Big[\left(\D+\D^T+\D^T\D+ \cancel{\D_d^T\D} + \cancel{\D^T\D_d}\right) + \\ 
-\left(\cancel{\D_d^T\D} + \D_d^T\D^T+\D_d^T\D^T\D+\D_d^T\D_d^T\D+ \cancel{\D_d^T\D^T\D_d}\right) +\\ 
-\left(\D\D_d + \cancel{\D^T\D_d} + \D^T\D\D_d + \cancel{\D_d^T\D\D_d} + \D^T\D_d\D_d\right) + \\ 
+ \left(\cancel{\D_d^T\D\D_d} + \cancel{\D_d^T\D^T\D_d} +\D_d^T\D^T\D\D_d+\D_d^T\D_d^T\D\D_d+\D_d^T\D^T\D_d\D_d\right)\Big].
\label{eq:expanded_E_Neumann}
\end{multline}
Finally, neglecting the terms $\mathcal{O}(\D_d^2)$, i.e. assuming that the first transformation $\Phi_1$ is linear, Eq. \eqref{eq:expanded_E_Neumann} reduces:
\begin{equation}
    \E_{2,N1} = \frac 12 \left(\D + \D^T + \D^T\D - \D_d^T\D^T - \D\D_d - \D_d^T\D^T\D - \D^T\D\D_d \right)
    \label{eq:E2Ns}
\end{equation}
The modified Green-Lagrange strain tensor $\E_{2,N1}$ is a polynomial function of the derivatives of the displacement fields $\ub$ and $\ub_d$, and can be thus used to compute a ROM using tensors.\\

\noindent
\textbf{Remark 1} \textit{(on Budiansky approximation).} The strain formulation in \cite{Budiansky1967}, used by Budiansky to study buckling in presence of defects, was obtained by subtracting the strain that a defect would produce on the nominal structure from the strain of the deformed structure measured with respect to the nominal configuration. It can be shown that truncating the Neummann series to the zero-th order (i.e. setting $N=0$, so that $\F_1^{-1}=\I$) and using Eq. \eqref{eq:E2} and \eqref{eq:F1expansion}, the strain writes:
\begin{equation}
    \E_{2,N0} = \frac 12 \left(\D + \D^T + \D^T\D + \D_d^T\D + \D^T\D_d\right),
    \label{eq:E2B}
\end{equation}
which is the same strain tensor we adopted in \citep{Marconi2020} following Budiansky's approximation.

\section{Finite Element formulation}
\label{sec:4}
In this section we derive the elastic internal forces (at element-level) for the FE discretization of the \textit{full order} model based on the strain as defined in Eq. \eqref{eq:E2Ns}. We remark that this full model represents just an approximation of the \textit{reference} full order model FOM-d (where the defect is embedded directly in the mesh by shifting the position of the nodes). Although not offering any direct advantage over FOM-d, this full model will allow us to compute the parametric ROM, as it will be explained in Section \ref{sec:S5}. 

First, it is convenient to switch to Voigt notation. Let $\thb=\{u_{,x}\,u_{,y}\,u_{,z}\,v_{,x}\,v_{,y}\,v_{,z}\,w_{,x}\,w_{,y}\,w_{,z}\}^T$ be the vectorized form of $\D$ and $\thb_d=\{u_{d,x}\,u_{d,y}\,u_{d,z}\,v_{d,x}\,v_{d,y}\,v_{d,z}\,w_{d,x}\,w_{d,y}\,w_{d,z}\}^T$ be the vectorized form of $\D_d$\footnote{For the derivatives of the displacement components, we use the notation $u_{,x}=\fpp{u}{x_0}$ and $u_{d,x}=\fpp{u_d}{x_0}$ (similar definition for $v$, $w$ and the other spatial coordinates).}. Eq. \eqref{eq:E2Ns} rewrites:
\begin{equation}
\E_{1,N1} = \left(\Hb + \frac 12 \A (\thb) + \Aii(\thb_d) + \Aiii(\thb_d) \A (\thb)\right) \thb
\label{eq:E1Ns}
\end{equation}
where
\begin{multline}
\textbf{H}= \small
\begin{bmatrix}
1 & 0 & 0 & 0 & 0 & 0 & 0 & 0 & 0 \\
0 & 0 & 0 & 0 & 1 & 0 & 0 & 0 & 0 \\
0 & 0 & 0 & 0 & 0 & 0 & 0 & 0 & 1 \\
0 & 1 & 0 & 1 & 0 & 0 & 0 & 0 & 0 \\
0 & 0 & 1 & 0 & 0 & 0 & 1 & 0 & 0 \\
0 & 0 & 0 & 0 & 0 & 1 & 0 & 1 & 0 \\
\end{bmatrix}, \normalsize
\hspace{.2cm}
\A(\thb)= \small
\begin{bmatrix}
u_{,x} & 0 & 0 & v_{,x} & 0 & 0 & w_{,x} & 0 & 0 \\
0 & u_{,y} & 0 & 0 & v_{,y} & 0 & 0 & w_{,y} & 0 \\ 
0 & 0 & u_{,z} & 0 & 0 & v_{,z} & 0 & 0 & w_{,z} \\
u_{,y} & u_{,x} & 0 & v_{,y} & v_{,x} & 0 & w_{,y} & w_{,x} & 0 \\
u_{,z} & 0 & u_{,x} & v_{,z} & 0 & v_{,x} & w_{,z} & 0 & w_{,x} \\
0 & u_{,z} & u_{,y} & 0 & v_{,z} & v_{,y} & 0 & w_{,z} & w_{,y} \\
\end{bmatrix},\\ \normalsize
\hspace{.2cm} \text{s.t.: } \hspace{.2cm} 
\D+\D^T+\D^T\D \longleftrightarrow (2\Hb+\A(\thb))\thb ,
\end{multline}
\begin{multline}
\Aii(\thb_d)= (-1)\small 
\begin{bmatrix}
u_{d,x} & v_{d,x} & w_{d,x} & 0 & 0 & 0 & 0 & 0 & 0\\
0 & 0 & 0 & u_{d,y} & v_{d,y} & w_{d,y} & 0 & 0 & 0\\
0 & 0 & 0 & 0 & 0 & 0 & u_{d,z} & v_{d,z} & w_{d,z}\\
u_{d,y} & v_{d,y} & w_{d,y} & u_{d,x} & v_{d,x} & w_{d,x} & 0 & 0 & 0\\
u_{d,z} & v_{d,z} & w_{d,z} & 0 & 0 & 0 & u_{d,x} & v_{d,x} & w_{d,x} \\
0 & 0 & 0 & u_{d,z} & v_{d,z} & w_{d,z} & u_{d,y} & v_{d,y} & w_{d,y} \\
\end{bmatrix},\\ \normalsize
\hspace{.2cm} \text{s.t.: } \hspace{.2cm} 
- \D^T_d\D^T - \D\D_d \longleftrightarrow 2\Aii(\thb_d)\thb ,
\end{multline}
\begin{multline}
\Aiii(\thb_d) = \left(-\frac 12 \right) \small 
\begin{bmatrix}
2u_{d,x} & 0 & 0 & v_{d,x} & w_{d,x} & 0 \\ 
0 & 2v_{d,y} & 0 & u_{d,y} & 0 & w_{d,y} \\ 
0 & 0 & 2w_{d,z} & 0 & u_{d,z} & v_{d,z} \\ 
2u_{d,y} & 2v_{d,x} & 0 & u_{d,x} + v_{d,y} & w_{d,y} & w_{d,x} \\ 
2u_{d,z} & 0 & 2w_{d,x} & v_{d,z} & u_{d,x} + w_{d,z} & v_{d,x} \\ 
0 & 2v_{d,z} & 2w_{d,y} & u_{d,z} & u_{d,y} & v_{d,y} + w_{d,z}
\end{bmatrix},
\normalsize \\
\hspace{.2cm} \text{s.t.: } \hspace{.2cm} 
- \D_d^T\D^T\D - \D^T\D\D_d \longleftrightarrow 2\Aiii(\thb_d)\A(\thb)\thb .
\end{multline}

Let us now define $\ub^e$ and $\ub_d^e$ as the nodal displacement vectors of a (continuum) finite element. Calling $\G$ the shape function derivatives matrix, such that $\thb=\G\ub^e$ and $\thb_d=\G\ub_d^e$, and exploiting the property by which $\A(\thb)\delta\thb=\A(\delta\thb)\thb$, the virtual variation of the strain in Eq. \eqref{eq:E1Ns} writes
\begin{equation}
    \delta\E_{1,N1} = \left(\Hb + \A + \Aii + 2 \Aiii \A \right)\G\delta\ub^e = \B_{N1}\delta\ub^e,
\end{equation}
where $\B$ is the strain-displacements matrix and where we dropped the explicit dependencies on $\thb_d$ and $\thb$ to ease the notation. The virtual work of internal forces is given by
\begin{equation}
    W_{int}= \int_{V_d} \delta\E_{1,N1}^T\textbf{S}_1 \text{ d}V_d =(\delta\ub^e)^T\int_{V_d}\B_{N1}^T\textbf{S}_1 \text{ d}V_d,
\end{equation}
where $\textbf{S}_1=\Cb\E_1$ is the Piola-Kirchhoff stress in Voigt notation, being $\Cb$ the linear elastic constitutive matrix, and where $V_d$ is the volume of the defected configuration. The expression for the internal forces $\f_{int}$ follows from the virtual work:
\begin{equation}
    \f_{int} = \int_{V_d}\B_{N1}^T\Cb\E_{1,N1} \text{ d}V_d.
    \label{eq:fint_BCE}
\end{equation}
Finally, the tangent stiffness matrix can be computed as usual taking the virtual variation of the internal forces (see \ref{appendixB}). Equations \eqref{eq:E1Ns} and \eqref{eq:fint_BCE} can be used to perform tests and/or simulations of the full model and to compare the results to the corresponding FOM-d in order to assess the quality of the approximation \textit{before} the reduction of the model.  In the next section, the DpROM derived from this formulation is presented.
\section{Tensorial representation of internal forces}
\label{sec:S5}
\subsection{Element--level tensors}
Under the hypothesis that for small defects $V_d \approx V_o$, Eq. \eqref{eq:fint_BCE} in full can be written as
\begin{equation}
    \f_{int} = \int_{V_o} \G^T\left(\Hb + \A + \Aii + 2 \Aiii \A \right)^T \Cb \left(\Hb + \frac 12 \A + \Aii + \Aiii \A \right) \G\ub^e \, \text{d} {V_o}.
    \label{eq:fint_full}
\end{equation}
To ``extract" the displacement vectors $\ub$ and $\ub_d$ from matrices $\A$, $\Aii$ and $\Aiii$, we can write:
\begin{subequations}
    \begin{align}
        \A &=\LL\cdot\thb = \LL\cdot (\G\ub^e), \\
        \Aii &= \textbf{L}_2\cdot\thb_d = \textbf{L}_2\cdot (\G\ub_d^e),\\
        \Aiii\A &= (\textbf{L}_3\cdot\thb_d) \cdot \thb = (\textbf{L}_3\cdot(\G\ub_d^e))\cdot(\G\ub^e),
    \end{align}
    \label{eq:Ls}
\end{subequations}
where $\LL,\textbf{L}_2 \in \mathbb{R}^{6 \times 9 \times 9}$ and $\textbf{L}_3 \in \mathbb{R}^{6 \times 9 \times 9 \times 9}$ are constant sparse matrices (see \ref{appendixA}).

Dropping for convenience the integral operation over the volume $V_o$ (implicitly assumed), we can separate the contributions in Eq. \eqref{eq:fint_full} as
\begin{subequations}
\begin{align}
    \f_1 &= \G^T \left( \Hb^T\Cb\Hb + \Hb^T\Cb\Aii + \Aii^T\Cb\Hb + \Aii^T\Cb\Aii \right) \G \ub^e, \\
    \begin{split}
        \f_2 &= \G^T \left( \frac 12 \Hb^T\Cb\A + \A^T\Cb\Hb +\frac 12 \Aii^T\Cb\A + \A^T\Cb\Aii + \right.\\ 
        & \hspace{1.36cm} 2\A^T\Aiii^T\Cb\Hb + \Hb^T\Cb\Aiii\A   + 2\A^T\Aiii^T\Cb\Aii + \Aii^T\Cb\Aiii\A \Big) \G \ub^e,
    \end{split}\\
    \f_3 &= \G^T \left( \frac 12 \A^T\Cb\A + \A^T\Cb\Aiii\A + \A^T\Aiii^T\Cb\A + 2\A^T\Aiii^T\Cb\Aiii\A \right) \G \ub^e,
\end{align}
\label{eq:f1f2f3}
\end{subequations}
where $\f_1$, $\f_2$ and $\f_3$ are the linear, quadratic and cubic terms in the displacement $\ub$, respectively.
These can be recasted in tensorial form as
\begin{subequations}
\begin{align}
    \f_1 &= \left[\tenb{K}{2n}{} + \tenb{K}{3d}{} \cdot \ub_d^e +\tenb{K}{4dd}{} : (\ub_d^e \otimes \ub_d^e)\right]\cdot\ub=\tenb{K}{2}{}(\ub_d^e) \cdot\ub^e, \\
    \f_2 &= \left[\tenb{K}{3n}{} + \tenb{K}{4d}{} \cdot \ub_d^e + \tenb{K}{5dd}{} : (\ub_d^e \otimes \ub_d^e) \right] : (\ub\otimes \ub )=\tenb{K}{3}{}(\ub_d^e) :(\ub^e \otimes \ub^e ), \\
    \f_3 &= \left[\tenb{K}{4n}{} + \tenb{K}{5d}{} \cdot \ub_d^e + \tenb{K}{6dd}{} : (\ub_d^e \otimes \ub_d^e) \right] \scalebox{.75}{ $\boldsymbol{\vdots}$ } (\ub \otimes \ub \otimes \ub)=\tenb{K}{4}{}(\ub_d^e)  \scalebox{.75}{ $\boldsymbol{\vdots}$ } (\ub^e \otimes \ub^e \otimes \ub^e).
\end{align}
\end{subequations}
\subsection{Reduced tensors and internal forces}
We now derive the reduced internal forces and tensors via Galerkin projection. Let us assume the following reduction for the displacement vectors $\ub^e$ and $\ub_d^e$:
\begin{subequations}
\begin{align}
    \ub^e &\approx \V^e \etab, \text{ with } \V^e \in\mathbb{R}^{n_e\times m},\, \etab\in\mathbb{R}^{m},   \\
    \ub_d^e &= \Vd^e \xib, \text{ with } \Vd^e \in\mathbb{R}^{n_e\times m_d},\, \xib\in\mathbb{R}^{m_d}\label{eq:Uxi}
\end{align}
\end{subequations}
being $\V^e$ and $\Vd^e$ the partitions of the RBs ($\V\in\mathbb{R}^{n\times m}$ and $\Vd\in\mathbb{R}^{n\times m_d}$, with $n$ number of dofs of the full order model) relative to the element, $\etab$ and $\xib$ the reduced coordinates. $m \ll n$ is thus the number of vectors included in the RB $\V$ while $m_d$ represents the number of the assumed shape defects, collected column-wise in $\Vd$. Introducing $\Gamma=\G\V^e$ and $\Upsilon=\G\Vd^e$, we can directly define the reduced order tensors using Einstein's notation as
\begin{subequations}
    \begin{align}
    \ten{Q}{2n}{IJ} &= \Gamma_{iI}  H_{ji} C_{jk} H_{kl} \Gamma_{kJ}, \\
    \ten{Q}{3d}{IJ\underline{K}} &= \Gamma_{iI}\left(H_{ji}C_{jk} \Lii_{kla} \Upsilon_{aK} + \Lii_{jia}\Upsilon_{aK} C_{jk}H_{kl}\right)\Gamma_{lJ},\\
    \ten{Q}{4dd}{IJ\underline{KL}} &= \Gamma_{iI}\Lii_{jia}\Upsilon_{aK} C_{jk} \Lii_{klb}\Upsilon_{bL}\Gamma_{lJ},\\
    \ten{Q}{3n}{IJK} &= \Gamma_{iI} \left( \frac 12 H_{ji}C_{jk} \Li_{kla} \Gamma_{aK} + \Li_{jia}\Gamma_{aK} C_{jk}H_{kl} \right) \Gamma_{lJ},\\
    \begin{split}
        \ten{Q}{4d}{IJK\underline{L}} &= \Gamma_{iI} \left( \frac 12 \Lii_{jia}\Upsilon_{aL} C_{jk} \Li_{klb}\Gamma_{bK} + \Li_{jia}\Gamma_{aK} C_{jk} \Lii_{klb}\Upsilon_{bL} \right. + \label{subeq:Q4d} \\
        & +  2\Liii_{jiab}\Upsilon_{bL}\Gamma_{aK} C_{jk} H_{kl} + H_{ji}C_{jk}\Liii_{klab}\Upsilon_{bL}\Gamma_{aK} \Big) \Gamma_{lJ},
    \end{split}\\
    \ten{Q}{5dd}{IJK\underline{LM}} &= \Gamma_{iI}\left( 2\Liii_{jiab}\Upsilon_{bL}\Gamma_{aK} C_{jk} \Lii_{klc}\Upsilon_{cM} + \Lii_{jia}\Upsilon_{aL} C_{jk} \Liii_{klbc}\Upsilon_{cM}\Gamma_{bK} \right) \Gamma_{lJ},\\
    \ten{Q}{4n}{IJKL} &= \frac 12\, \Gamma_{iI}\Li_{jia}\Gamma_{aK}C_{jk}\Li_{klb}\Gamma_{bL}\Gamma_{lJ}, \\ 
    \ten{Q}{5d}{IJKL\underline{M}} &= \Gamma_{iI}\left(\Li_{jia}\Gamma_{aK}C_{jk}\Liii_{klbc}\Upsilon_{cM}\Gamma_{bL} + \Liii_{jiab}\Upsilon_{bM}\Gamma_{aK}C_{jk}\Li_{klc}\Gamma_{lL}\right)\Gamma_{lJ}, \\
    \ten{Q}{6dd}{IJKL \underline{MN}} &= 2\, \Gamma_{iI}\Liii_{jiab}\Upsilon_{b M}\Gamma_{a L} C_{jk} \Liii_{klcd}\Upsilon_{d N} \Gamma_{c K} \Gamma_{lJ}.
    \end{align}
    \label{eq:reduced_tensors}
\end{subequations}
where, for convenience, tensor dimensions of size $m$ are denoted by capital letter subscripts, dimensions of size $m_d$ by underlined capital letter ones. So, for example, $\tenb{Q}{3d}{} \in \mathbb{R}^{m \times m \times m_d}$. The global reduced tensors of the full structure can then be computed directly summing up the element contributions, a procedure which is highly parallelizable. Reduced (global) internal forces can therefore be defined as $\f_{int,r} = \f_{1r} + \f_{2r} + \f_{3r}$, where
\begin{subequations}
\begin{align}
    \f_{1r} &= \left[\tenb{Q}{2n}{} + \tenb{Q}{3d}{} \cdot \xib +\tenb{Q}{4dd}{} : (\xib \otimes \xib)\right]\cdot\etab=\tenb{Q}{2}{}(\xib) \cdot\etab, \\
    \f_{2r} &= \left[\tenb{Q}{3n}{} + \tenb{Q}{4d}{} \cdot \xib + \tenb{Q}{5dd}{} : (\xib \otimes \xib) \right] : (\etab\otimes \etab )=\tenb{Q}{3}{}(\xib) :(\etab\otimes \etab ), \\
    \f_{3r} &= \left[\tenb{Q}{4n}{} + \tenb{Q}{5d}{} \cdot \xib + \tenb{Q}{6dd}{} : (\xib \otimes \xib) \right] \scalebox{.75}{ $\boldsymbol{\vdots}$ } (\etab \otimes \etab \otimes \etab)=\tenb{Q}{4}{}(\xib)  \scalebox{.75}{ $\boldsymbol{\vdots}$ } (\etab \otimes \etab \otimes \etab),
\end{align}
\label{eq:reduced_int_forces}
\end{subequations}
and the reduced tangent stiffness matrix $\tenb{Q}{t}{}$ can be written simply as
\begin{equation}
\ten{Q}{t}{IJ} = \ten{Q}{2}{IJ} + (\ten{Q}{3}{IJj} + \ten{Q}{3}{IjJ})\eta_j + (\ten{Q}{4}{IJij} + \ten{Q}{4}{IiJj} + \ten{Q}{4}{IijJ})\eta_i\eta_j.
\end{equation}

Finally, the equations of motion for the reduced system write:
\begin{equation}
\textbf{M}_r \ddot\etab(t) + \textbf{C}_r \dot\etab(t) + \f_{int,r}(\etab(t),\xib) = \f_{ext,r}(t)
\end{equation}
where $\textbf{M}_r = \V^T \textbf{M} \V$ and $\textbf{C}_r = \V^T \textbf{C}_d \V$ are the reduced mass and damping matrices, $\f_{ext,r}(t)=\V^T \f_{ext}(t)$ the reduced external forces acting on the system. Notice that, in accordance to the hypothesis made for the internal forces, also these matrices must be integrated over the nominal volume $V_o$.\\

\noindent
\textbf{Remark 2} \textit{(on tensor computation)}. Equations \eqref{eq:reduced_tensors} give directly the stiffness tensors in reduced form, and this is in general highly desirable as their integration over the element volume takes multiple evaluations (e.g. through Gauss quadrature). Since the computational complexity highly depends on the number of dofs of the tensor, it is preferable to integrate directly the reduced ones as long as the number of reduced coordinates $m$ is lower than the number of element's dofs $n_e$ (e.g. $n_e=60$ for a serendipity hexahedron with quadratic shape functions). In case $m>n_e$, it is computationally more efficient to compute the element tensors first (using Eqs \eqref{eq:reduced_tensors} and replacing both $\Gamma$ and $\Upsilon$ with $\G$) for Gauss integration, then project the element tensors using $\V$ and $\Vd$ accordingly. A similar reasoning can be done for $m_d$, but under the very likely hypothesis that $m_d \ll n_e$ it results almost always convenient to adopt the reduced form.

\subsection{Volume integration}
\label{sec:5.3}
The tensors in Eqs. \eqref{eq:reduced_int_forces} can be computed once and for all integrating on the nominal volume $V_o$, under the said hypothesis that the defect does not change the defected volume $V_d$ significantly. When this hypothesis cannot be made, one can adopt the following approximation. Let $\textbf{Q}_e$ be the generic expression of a tensor to be integrated over the volume $V_d$ (element level). We can compute the final reduced tensor \textbf{Q} as:
\begin{equation}
\textbf{Q} = \sum_{e=1}^{N_{el}}\int_{V_d} \textbf{Q}_e \text{ d}V_d = \sum_{e=1}^{N_{el}}\int_{V_o} \textbf{Q}_e \text{det}(\F_1) \text{ d}V_o.
\label{eq:correct_integral}
\end{equation}
where $N_{el}$ is the total number of elements. The determinant of $\F_1$ can now be approximated retaining only first order terms. To the purpose of illustration, let us consider the following 2D example where the global defect is given by the linear superposition of two shape defects, that is:
\begin{equation}
\ub_d(\xb_0, \xib) = \begin{Bmatrix}
u_d \\
v_d
\end{Bmatrix} = 
\begin{bmatrix}
f_u^{(1)} (\xb_0) & f_u^{(2)} (\xb_0)\\
f_v^{(1)} (\xb_0) & f_v^{(2)} (\xb_0)
\end{bmatrix}
\begin{Bmatrix}
\xi_1 \\
\xi_2
\end{Bmatrix}
=\left[\f^{\,(1)},\, \f^{\,(2)} \right]\xib.
\end{equation}
where we denote with $\f^{\,(i)}=[f_u^{(i)},\,f_v^{(i)}]^T$ the vector of the functions describing the i-th shape-defect for the x-displacement $u_d$ and the y-displacement $v_d$, respectively. We can approximate the determinant of $\F_1$ as
\begin{align*}
\text{det}(\F_1) =& \,1 + u_{d,x} + v_{d,y} + u_{d,x}v_{d,y} - u_{d,y}v_{d,x} \\
=& \,1 + \xi_1 \left(f_{u,x}^{(1)} + f_{v,y}^{(1)} \right) + \xi_2 \left(f_{u,x}^{(2)} + f_{v,y}^{(2)} \right) + \xi_1^2 \left( f_{u,x}^{(1)} f_{v,y}^{(1)} - f_{u,y}^{(1)} f_{v,x}^{(1)} \right) + \\
& + \xi_1\xi_2 \left( f_{u,x}^{(1)} f_{v,y}^{(2)} + f_{u,x}^{(2)} f_{v,y}^{(1)} - f_{u,y}^{(1)} f_{v,x}^{(2)} - f_{u,y}^{(2)} f_{v,x}^{(1)} \right) + \xi_2^2 \left(f_{u,x}^{(2)} f_{v,y}^{(2)} - f_{u,y}^{(2)} f_{v,x}^{(2)} \right) \\
\approx& \,1 + \xi_1 \left(f_{u,x}^{(1)} + f_{v,y}^{(1)} \right) + \xi_2 \left(f_{u,x}^{(2)} + f_{v,y}^{(2)} \right)
\end{align*}
where we neglected higher order terms. Generalizing this result for $m_d$ defects, we can write
\begin{equation}
\text{det}(\F_1) \approx 1 + \sum_{i=1}^{m_d} \xi_i \left( \text{div } \f^{\,(i)} \right)
\end{equation}
so that Eq. \eqref{eq:correct_integral} can be approximated as:
\begin{equation}
\textbf{Q} \approx \sum_{e=1}^{N_{el}}\int_{V_o} \textbf{Q}_e \text{ d}V_0 + \sum_{i=1}^{m_d} \left(\xi_i \sum_{e=1}^{N_{el}}\int_{V_o} \textbf{Q}_e \left(\text{div } \f^{\,(i)} \right) \text{ d}V_o \right) = \textbf{Q}^\prime + \sum_{i=1}^{m_d}\xi_i\textbf{Q}^{\prime\prime}_i
\label{eq:tensor_volume_correction}
\end{equation}
where $\textbf{Q}^\prime$ is the tensor evaluated on the nominal volume and $\textbf{Q}^{\prime\prime}_i$ is the  contribution of the i-th defect, which can be computed again \textit{once and for all} offline, referring to the nominal volume. The additional computational burden to compute $\textbf{Q}^{\prime\prime}_i$ grows less than linearly with the number of defects, since in a quadrature integration scheme we can use the $\textbf{Q}_e$ evaluated at integration points (using Eqs. \eqref{eq:reduced_tensors}) both for $\textbf{Q}^{\prime}_i$ and for $\textbf{Q}^{\prime\prime}_i$. The additional computations therefore involve only scalar by tensor multiplications and tensor sums, so that most of the added computational time is merely due to memory access management. Notice that one could also compute all the additional tensors needed to describe det($\F_1$) with no approximation (even tough this is in most cases unnecessary, for h.o.t. do not improve accuracy significantly). However, the first order approximation we presented has the advantage to introduce only one new term for every additional defect. Finally, if adopting this correction one should use the mass matrix computed over the defected volume $V_d$ in accordance with the present formulation. This way a new mass matrix $\textbf{M}_d$ must be computed for each new parameter realization, however, being the matrix constant during the analysis, this additional cost is negligible.\\

\noindent
\textbf{Remark 3} \textit{(on computational efficiency)}. The corrective terms $\textbf{Q}^{\prime\prime}_i$ in Eq. \eqref{eq:tensor_volume_correction} are null for an isochoric transformation between nominal and defected domain (det($\F_1$)=1). In practice, one can set up a procedure to avoid the computation of these terms to speed up the construction of the reduced tensors. 

\section{Reduction Basis}
\label{sec:6}
To construct the system described so far it is necessary to select the bases $\V$ and $\Vd$. The latter is simply a collection of \textit{user-defined} displacement vectors, each representing one specific defect, so that the final defected shape is given by a linear superposition (see Eq. \eqref{eq:Uxi}). The (properly said) RB is $\V$, whose choice may not be trivial, as it must correctly represent the system response over a range of parameters without the possibility to be changed (since a change would require to recompute the stiffness tensors). As previously done in \citep{Marconi2020}, our choice is to use a modal-based approach including VMs, MDs and Defect-Sensitivities (DSs) \cite{Hay2010} in the RB, as this solution offers a way to construct a basis in a direct way, that is without convoluted basis selection strategies, the need of computing all (or an excessively high number of) eigenvectors or the need for previously computed simulations. We remark however that, in principle, one could use also other RBs, as long as they are valid over the parameter space.

Let us consider the following eigenvalue problem
\begin{equation}
    \left( \textbf{K}_t - \omega_i^2\textbf{M} \right) \Phib_i = \textbf{0}
    \label{eq:EVP}
\end{equation}
where $\textbf{K}_t=\textbf{K}_t(\ub,\ub_d)$ is the tangent stiffness matrix, \textbf{M} the mass matrix, $\omega_i$ the i-th eigenfrequency and $\Phib_i$ the corresponding eigenvector. Static Modal Derivatives $\thb_{ij}$ (MDs) are computed neglecting the mass term, by taking the derivative of Eq. \eqref{eq:EVP} with respect to $\eta_j$ and evaluating the resulting expression at equilibrium (i.e. $\eta_j=0$) and for $\xib=\textbf{0}$:
\begin{equation}
    \thb_{ij} = \frac{\partial \Phib_i}{\partial \eta_j} \bigg|_{0} = -\textbf{K}_{0}^{-1}\frac{\partial \textbf{K}_t(\Phib_j\eta_j,\textbf{0})}{\partial \eta_j} \bigg|_{0} \Phib_i.
    \label{eq:MD}
\end{equation}

Defect Sensitivities (DSs) $\Xib_{i,j}$ can be obtained following a similar procedure, differentiating with respect to $\xi_j$:
\begin{equation}
    \Xib_{i,j} = \frac{\partial \Phib_i}{\partial \xi_j}\bigg|_{0} = -\textbf{K}_{0}^{-1}\frac{\partial \textbf{K}_t(\textbf{0},\Vd_j\xi_j)}{\partial \xi_j} \bigg|_{0} \Phib_i.
    \label{eq:DS}
\end{equation}

Expressions for the tangent stiffness derivatives are given in \ref{appendixB}.\\

%
%
%

\begin{table}[]
\centering
\caption{Acronyms for the different models considered in the numerical studies.}
\label{tab:acronyms}
\resizebox{\textwidth}{!}{%
\begin{tabular}{ll}
\hline
Model     & Description                                                                                                 \\ \hline
\textbf{FOM-d}     & Full Order Model with defect included by shifting the mesh nodes from the nominal configuration \\
                   & (no approximation). It is the reference model.         \\
\textbf{ROM-d}     & Reduced Order Model computed from FOM-d. Its reduction basis comprises VMs and MDs.                         \\
\textbf{DpROM-N0}  & Defect parametric Reduced Order Model, based on the 0-th order Neumann expansion (see Eq. \eqref{eq:E2B}).  \\
          & Its reduction basis comprises VMs, MDs and DSs. Tensors are up to the 4-th order (see \citep{Marconi2020}).                                                           \\
\textbf{DpROM-N1}  & Defect parametric Reduced Order Model, based on the 1-st order Neumann expansion (see Eq. \eqref{eq:E2Ns}). \\
          & Its reduction basis comprises VMs, MDs and DSs. Tensors are up to the 6-th order (see Eqs. \eqref{eq:reduced_tensors}).                                                          \\
\textbf{DpROM-N1t} & Truncated version of DpROM-N1, where terms of order $\mathcal{O}(\D_d\D^2)$ in Eq. \eqref{eq:E2Ns} have been neglected.         \\
          & As a consequence, $\tenb{Q}{5d}{}$, $\tenb{Q}{5dd}{}$, $\tenb{Q}{6dd}{}$ and the last two terms in Eq. \eqref{subeq:Q4d} are null.                                                     \\
\textbf{-v} (suffix) & as the corresponding DpROM, but with the volume integration correction described in Section \ref{sec:5.3}.                            \\ \hline
\end{tabular}%
}
\end{table}

\noindent
\textbf{Remark 4} \textit{(on higher derivatives)}. Given the higher accuracy of the model, larger defect magnitudes can be considered as compared to \cite{Marconi2020}. To fully exploit the increased applicability range, a richer RB might be necessary, reason why we here introduce second Defect-Sensitivities (DS2s) and MDs-Sensitivities (MDSs). Let us take the derivative of Eq. \eqref{eq:MD} with respect to the k-th defect amplitude $\xi_k$. We define the MDS $\thb_{ij,k}$ as:
\begin{equation}
    \thb_{ij,k} = \frac{\partial \thb_{ij}}{\partial \xi_k}\bigg|_{0} = -\textbf{K}_{0}^{-1} \left( \frac{\partial \textbf{K}_t}{\partial \xi_k} \bigg|_{0} \thb_{ij} + \frac{\partial^2 \textbf{K}_t}{\partial \eta_j \partial \xi_k} \bigg|_{0} \Phib_i + \frac{\partial \textbf{K}_t}{\partial \eta_j} \bigg|_{0} \Xib_{i,k} \right).
    \label{eq:MDS}
\end{equation}
Notice that $\thb_{ij,k} \neq \thb_{ji,k}$. In the same manner, the second Defect Sensitivities (DS2s) with respect to $\xi_k$ write:
\begin{equation}
    \Xib_{i,jk} = \frac{\partial \Xib_{i,j}}{\partial \xi_k}\bigg|_{0} = -\textbf{K}_{0}^{-1} \left( \frac{\partial \textbf{K}_t}{\partial \xi_k} \bigg|_{0} \Xib_{i,j} + \frac{\partial^2 \textbf{K}_t}{\partial \xi_j \partial \xi_k} \bigg|_{0} \Phib_i + \frac{\partial \textbf{K}_t}{\partial \xi_j} \bigg|_{0} \Xib_{i,k} \right).
    \label{eq:DS2}
\end{equation}

It is evident that the blind inclusion of DS2s and/or MDSs in the RB would add an unacceptable number of unknowns, especially when considering MDSs. Depending on the type of the analysis (linear/nonlinear), on the kind of the defect (i.e. affecting the linear or the nonlinear dynamics) and on the entity of the defect itself (large/small), one can decide whether to include some vectors or not. Pre-selection strategies to reduce the basis size, as the one presented in \cite{Tiso2011} and \cite{Jain2017}, are beyond the scopes of this work and are not treated hereafter.

\section{Numerical tests -- I}
\label{sec:7}

\begin{figure}[t!]
  \centering
  \includegraphics[width=.99\linewidth]{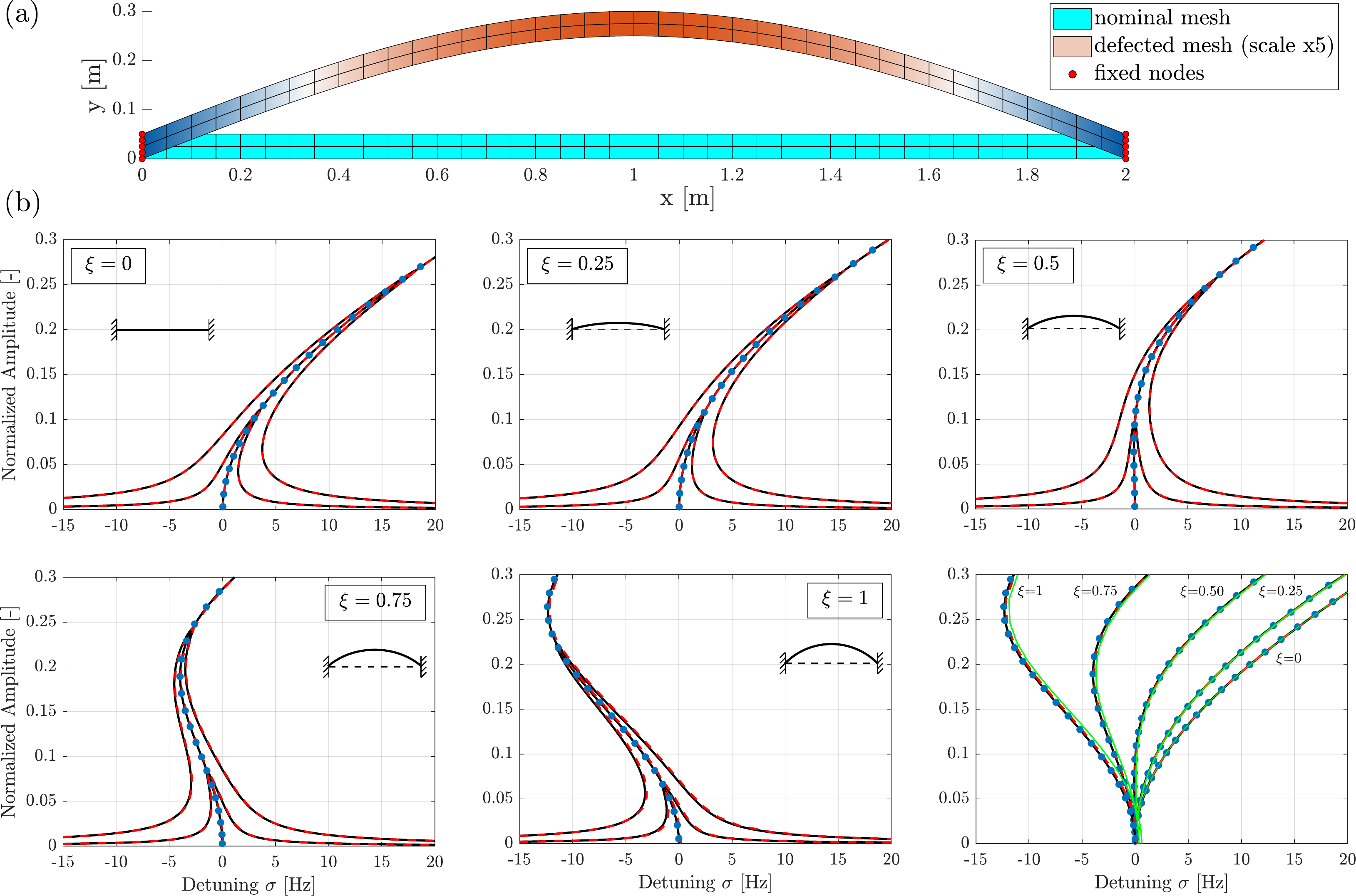}
  \caption{(a) Model-I: Nominal mesh and defected mesh with $\xi=1$ (and a x5 scale factor). (b) Frequency Responses and backbone curves for different defect amplitudes $\xi$ using the Harmonic Balance method (7 harmonics) for: ROM-d ({\hwplotA}), DpROM-N1 ({\hwplotB}), DpROM-N0 ({\hwplotC}, only backbones). Backbones have been computed also for the FOM-d ({\hwplotD}) using the \textit{shooting method} for validation. The vertical displacement of the mid-span of the beam is shown (first harmonic coefficient of the Fourier series, normalized over the beam thickness). For each plot, the detuning parameter $\sigma$ is referred to the corresponding FOM-d first eigenfrequency. The bottom-right figure collects the backbone curves for comparison.}
  \label{fig:beam_FR}
\end{figure}

We consider now a FE model of an aluminum beam, of length $l_x=2\,m$, thickness $t_y=50\,mm$ and width $w_z=0.2\,m$, clamped at both ends. We use a 2D plain strain model, with a mesh of 80 quadrilateral elements with quadratic shape functions (630 dofs). A Rayleigh damping matrix $\textbf{C}_d=\alpha \textbf{M}_d + \beta \textbf{K}_0$ is introduced by imposing a quality factor $Q_1=Q_2=100$ on the first and second modes of the linear system ($\alpha=3.1,\,\beta=6.3\cdot10^{-6}$). A nodal load $F$ is applied to the mid-span of the beam (with $F=1$ kN and $F=4$ kN for the forced responses). A shape defect defined as the vertical translation of the nodes $v_d(x,\xi) = \xi t_y \sin( \pi/l_x x)$ is imposed, deforming the nominal geometry of the straight beam into a shallow arch. Notice that this kind of defect represents an isochoric transformation (see Section \ref{sec:5.3}).

Table \ref{tab:acronyms} reports the acronyms used for the models of this and the next numerical studies. For $\xi = \{0,\,0.25,\,0.5,\,0.75,\,1\}$, backbones and Frequency Responses (FR) are computed for ROM-d, DpROM-N1 and DpROM-N0, constructed using 5 VMs, 15 MDs, and 5 DSs (only for DpROMs). The Harmonic Balance (HB) method was used (with 7 harmonics) using the NLvib Matlab tool \cite{Krack2019} (slightly modified to adapt the direct use of tensors) and our in-house Matlab FE code. To validate the results of the ROMs, the Shooting Method is used to find the backbones of the corresponding FOM-d. Results are shown in Fig. \ref{fig:beam_FR}. Computations were carried out in Matlab 2020a on a local machine equipped with an Intel(R) Xeon(R) Silver 4214 CPU @2.20 GHz and 256 GB RAM @2666 MHz. Tensors were built in a Julia sub-routine, called by the main Matlab code, which uses the TensorOperations package \cite{jutho_2019_3245497} for the tensor contraction. At present, the tensor construction is implemented serially, therefore leaving space for possible future speed-ups exploiting parallel computing, as remarked earlier. 

\begin{table}[t!]
\centering
\caption{Average computational times. The data of the two DpROMs, being very similar, are clustered togeter. For the FR, the times refer to the higher forcing ($F=4$ kN). ROM-d, DpROM-N1/0 and FOM-d have 20, 25 (due to the 5 additional DSs) and 630 dofs, respectively. Notice that, being the size of the FOM-d very small, no significant conclusions in terms of speedups can be drawn from this data (refer to the next section for a detailed discussion).}
\label{tab:comptimes_FRF}
\begin{tabular}{@{}llccc@{}}
\toprule
Model--I           	      &                             & \textbf{ROM-d} & \textbf{DpROM-N1/0} & \textbf{FOM-d} \\ \midrule
\textbf{Harmonic Balance (HB)} & \textit{Frequency Response} & 649 s          & 673 s               & --             \\
                          & \textit{Backbone}           & 237 s          & 273 s               & --             \\
\textbf{Shooting}         & \textit{Backbone}           &  31 min        &  35 min             & 83 h 18 min    \\
\textbf{ROM construction} & \multicolumn{1}{c}{}        & 0.97 s         & 6.5 s / 2 s         & --             \\ \bottomrule
\end{tabular}
\end{table}

As it can be observed, the shift from hardening to softening behavior is well captured by all the models, with a minor loss of accuracy of the DpROMs as $\xi$ increases. In particular, DpROM-N0 shows a significant frequency offset of the first linear eigenfrequency which remains constant throughout the backbone curve (the same happens for the FRs, but we omitted to plot them for the sake of figures clarity). 
The main goal of the present test was to assess the accuracy of the method verifying the results against the FOM and over a range of frequencies. However, computational times are collected in Table \ref{tab:comptimes_FRF} for completeness. Run times for the shooting method with the (Dp)ROMs are included for comparison. These figures, however, must be taken just as an indication, first because of the difference between FOM and ROMs in terms of convergence during continuation (ROMs are less likely to incur into numerical artifacts) and, secondly, because speed and convergence of this kind of analysis is highly sensitive to several parameters and finding the best combination by trial-and-error usually leads to sub-optimal performances. Last but not least, the size of the FOM in this case is too low to really appreciate the savings in terms of ROM construction.

\section{Numerical tests -- II}
\label{sec:8}

\subsection{MEMS gyroscope}
\begin{figure}[t]
\begin{subfigure}{.5\textwidth}
  \centering
  \includegraphics[width=.99\linewidth]{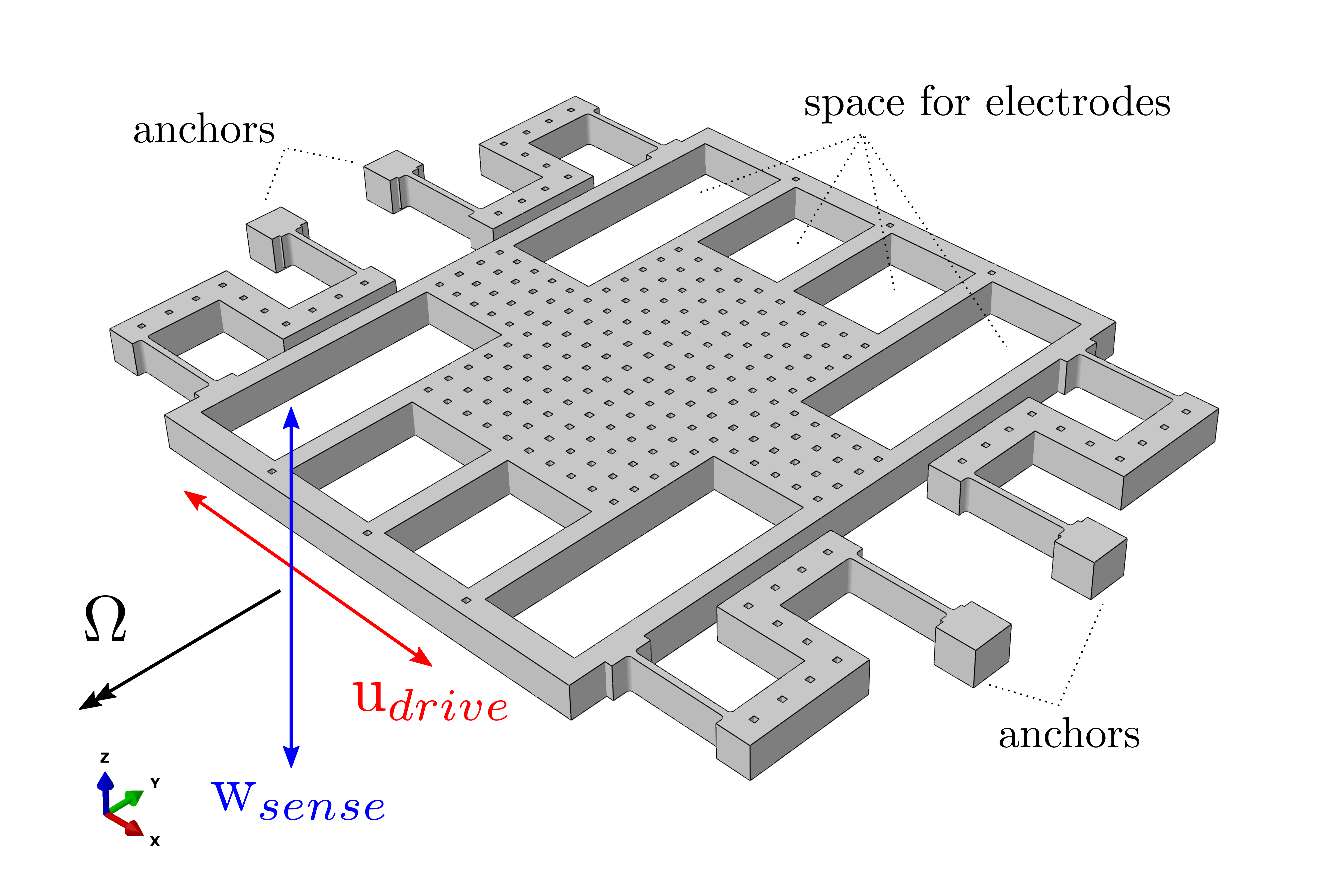}
  \caption{}
  \label{subfig:figura_TS1_model}
\end{subfigure}
\begin{subfigure}{.5\textwidth}
  \centering
  \includegraphics[width=.99\linewidth]{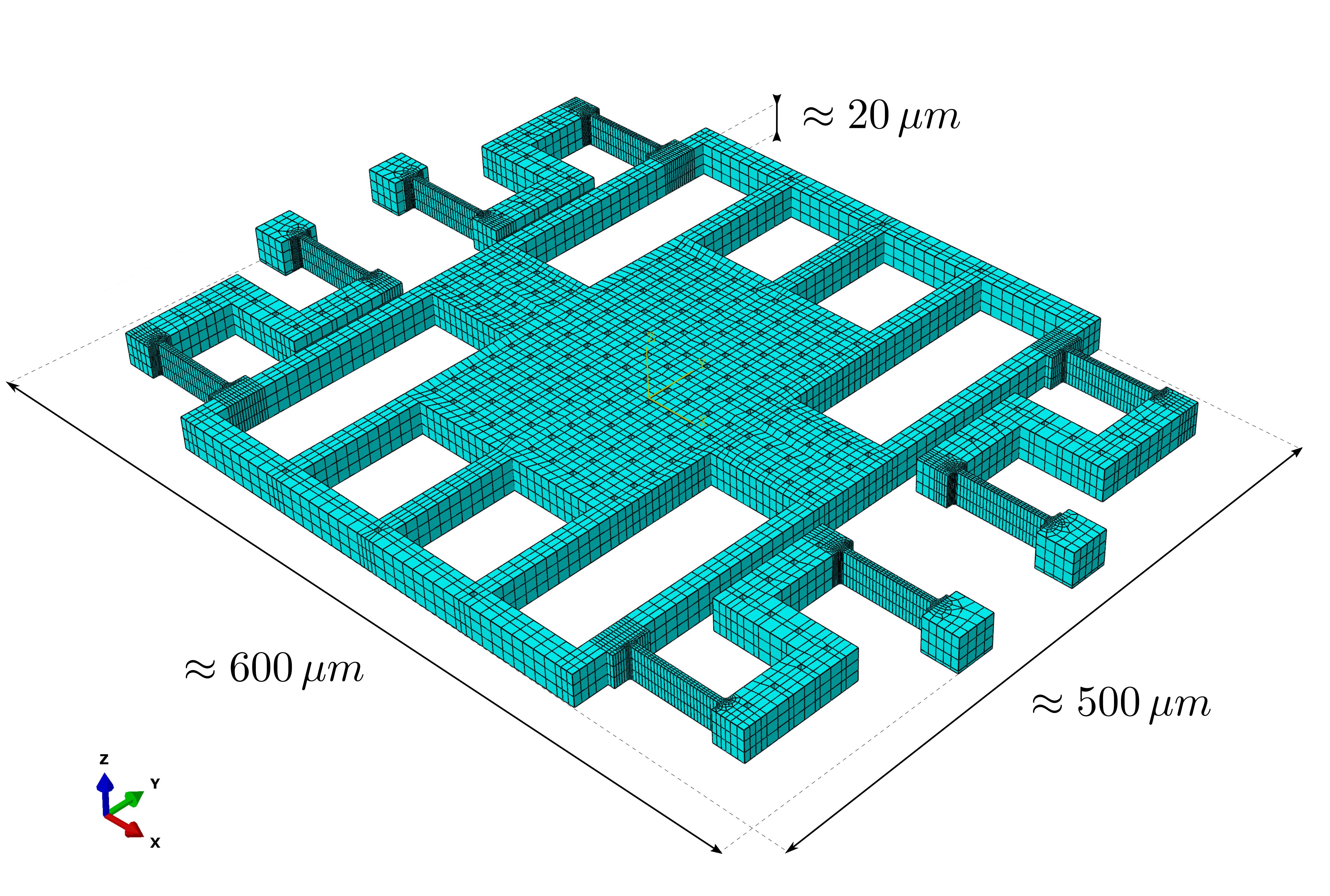}
  \caption{}
  \label{subfig:figura_TS1_mesh}
\end{subfigure}
\caption{(a) Model--II, a MEMS gyroscope. Drive and sense direction are indicated by arrows. (b) Meshed model, with 14,920 quadratic hexahedra, for a total of 87,767 nodes and 261,495 dofs. Approximate dimension are given.}
\label{fig:figura_TS1}
\end{figure}

The last example we present is a prototype MEMS mono-axial gyroscope, shown in Fig. \ref{subfig:figura_TS1_model}. The device consists in a mass suspended by four S-shaped springs, connected to the ground on the bottom of the anchors. It is a monolithic piece, produced via Deep Reactive Ion Etching (DRIE), a process which removes material from a plane silicon wafer to obtain the final geometry. The etching procedure is the main cause of production shape defects of MEMS devices, as it will be detailed later. In operative conditions, the mass is kept in motion by comb finger electrodes at the natural frequency of the drive mode (i.e. a mode featuring motion mainly in the x-direction), so that in presence of an external angular rate $\Omega$ (along the y-axis) a vertical displacement $w_{sense}$ arises due to Coriolis effect along the z-axis (sense). The latter is then converted into an electrical signal through the parallel plate electrode placed on the ground below the mass, providing the measure for the angular rate. In general, a defect or a combination of them may create a coupling between the x- and z-axes so that the drive motion generates an additional out-of-plane displacement which superimposes to the Coriolis displacement to be measured. This is usually referred to as \textit{quadrature error} since, being proportional to the drive displacement, it is in phase quadrature with the Coriolis signal, proportional to the drive velocity. Though it is possible to tell apart the two contributions, this is highly undesirable as it requires dedicated, over-sized electronics to accommodate the larger displacements. Ultimately, this results in higher power consumption.

\subsection{FE model, defects and simulation details}
\begin{figure}[t!]
\centering
\begin{subfigure}{.45\textwidth}
  \centering \includegraphics[width=.99\linewidth]{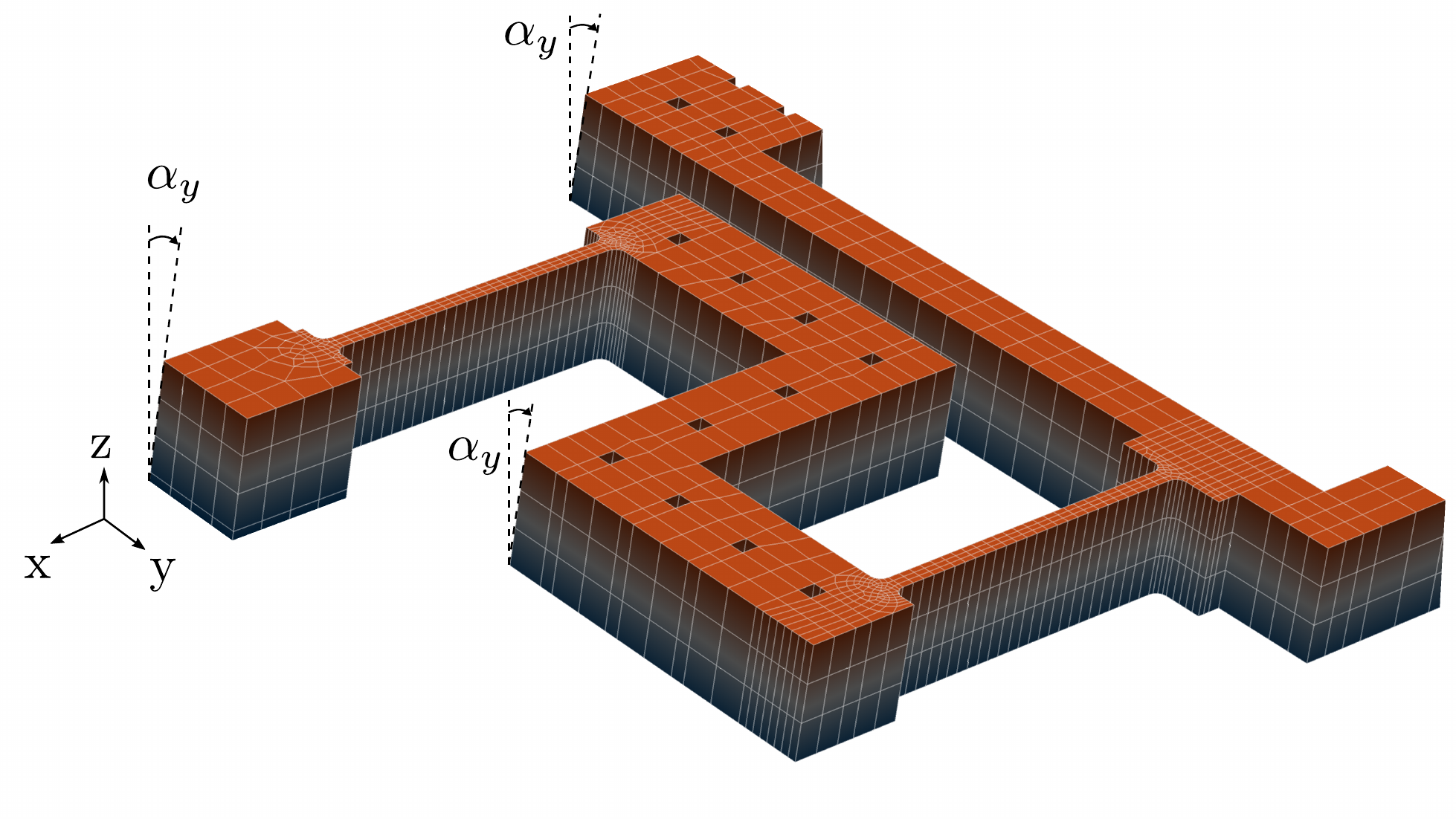}
  \caption{}
  \label{subfig:figura_ts1_def1}
\end{subfigure}
\begin{subfigure}{.45\textwidth}
  \centering \includegraphics[width=.99\linewidth]{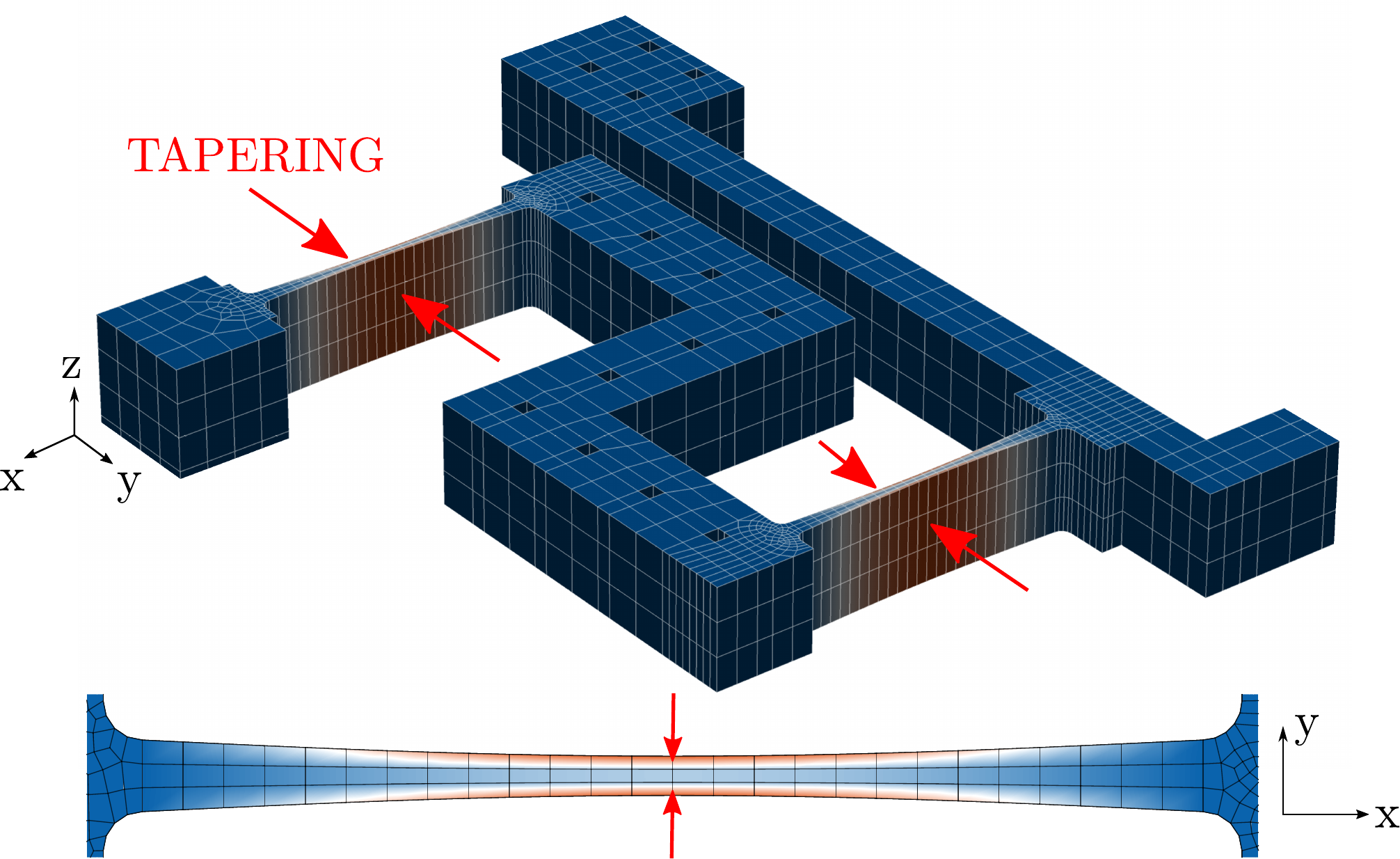}
  \caption{}
  \label{subfig:figura_ts1_def2}
\end{subfigure}
\caption{(a) First defect shape, $\Vd_1$: constant wall angle $\alpha_y$ (only one beam is shown). The colormap indicates x-displacement. (b) Second defect shape, $\Vd_2$: tapering of the suspension beams, 3D and top views (only one beam is shown). The colormap indicates y-displacement (absolute value).}
\label{fig:figura_TS1_def}
\end{figure}

The FE model is shown in Fig. \ref{subfig:figura_TS1_mesh} and describes in detail the geometry and mesh of the device, counting 14,920 quadratic hexahedral elements for a total of 261,495 dofs. For the present study we selected two typical defects occurring in the production of MEMS devices, namely the \textit{wall angle} (shown in Fig. \ref{subfig:figura_ts1_def1}) and a restriction of the cross section of the beams (Fig. \ref{subfig:figura_ts1_def2}). The first is generated by the fact that the plasma beam of the DRIE process might be not perfectly perpendicular to the working plane, while the second one typically comes from an overexposure to the chemical attacks (\textit{over-etching}). In the spirit of our method, we can describe the global defects as the superposition of these two displacement fields (see Eq. \eqref{eq:Uxi}), letting $\Vd = [\Vd_1,\,\Vd_2]$ with the associated amplitude parameter vector $\xib=[\xi_1,\,\xi_2]^T$. The wall angle shape defect $\Vd_1=[\ub_{d1},\,\textbf{v}_{d1},\,\textbf{w}_{d1}]^T$ is given by 
\begin{equation}
u_{d1}(\xi_1,z)=\xi_1\tan(\alpha_y)z,
\label{eq:wall_angle}
\end{equation}
and $v_{d1}=w_{d1}=0$. The tapering of the beams $\Vd_2=[\ub_{d2},\,\textbf{v}_{d2},\,\textbf{w}_{d2}]^T$ is defined as 
\begin{equation}
v_{d2}(\xi_2,x,y) = \xi_2\sin\left( \frac{\pi}{L_b} (x-x_{off})\right)(y-y_{mid})\frac{2}{W_b},
\end{equation}
and $u_{d2}=w_{d2}=0$, where $L_b$ and $W_b$ are the length and the width of the beam, $x_{off}$  an offset depending on the location of each beam and $y_{mid}$ is the y-coordinate corresponding to the middle line of each beam. To ease the interpretation of the amplitude parameters, in the following $\xi_1$ is reported in degrees to represent the physical wall angle coming from the product $\xi_1 \tan (\alpha_y)$ in Eq. \eqref{eq:wall_angle}, while $\xi_2$ is reported as a percentage of the beam thickness.

We compute the FR of the MEMS gyroscope using the NLvib Matlab tool and our in-house Matlab FE code. We used a reduction basis with 3 VMs, the corresponding 6 MDs and 3 DSs per defect (only for the DpROMs), and we used the HB method with $H=5$ harmonics (with $N_s=3H+1$ time samples per period, the minimum number of samples by which no sampling error is introduced in the harmonics up the the H-th order when considering polynomial nonlinearities up to the third order \cite{woiwode:hal-02424746}, as in our case). Given the size of the model, we take as reference the results of ROM-d, as it would be prohibitively time and memory consuming to compute the frequency response for FOM-d. Apart from the practical issues, we justify this choice considering on the one side the good results obtained for lower dimensional models (as the one presented in the previous section), and on the other side considering that, ultimately, our DpROMs will be at best as good as ROM-d, which is not parametric and not approximated in its formulation.

\begin{figure}[t!]
\centering
\begin{subfigure}{.99\textwidth}
  \centering
  \includegraphics[width=1\linewidth]{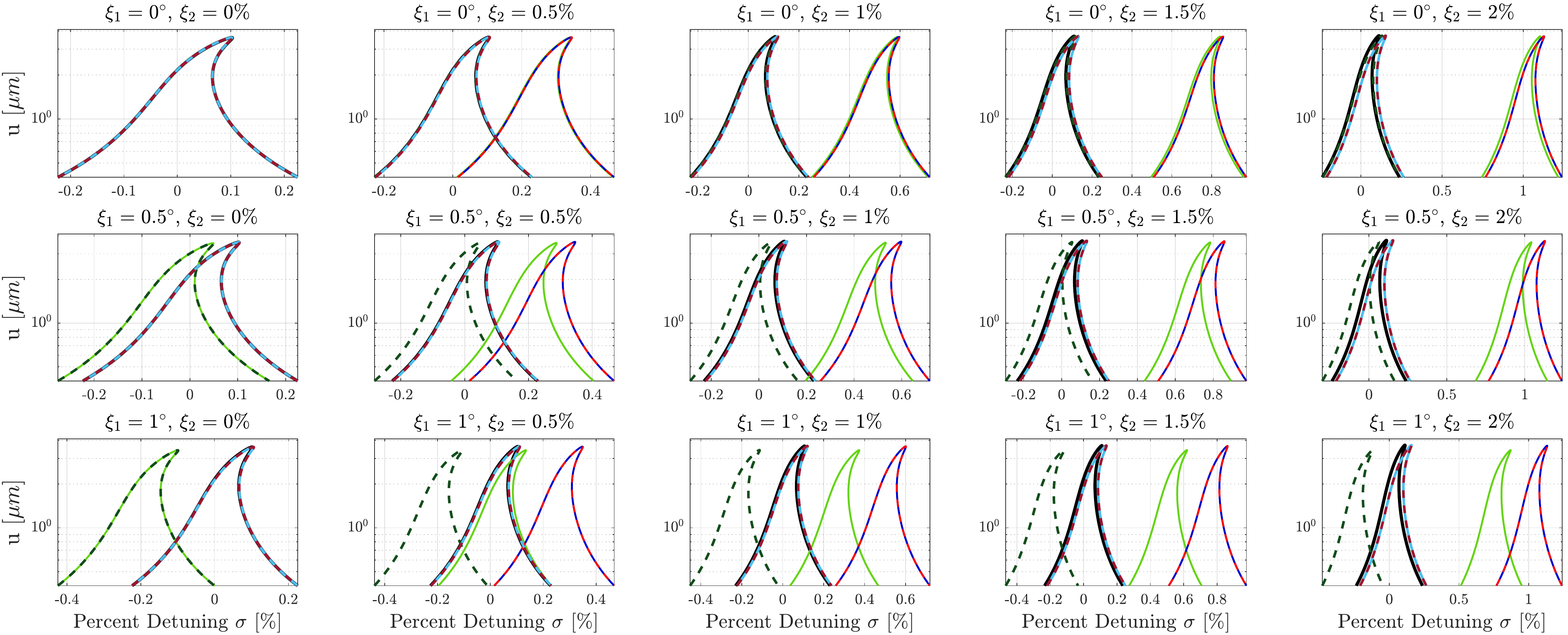} 
  \caption{Drive response.}
  \label{subfig:figura_TS1_u}
\end{subfigure}\\ \vspace{5mm}
\begin{subfigure}{.99\textwidth}
  \centering
  \includegraphics[width=1\linewidth]{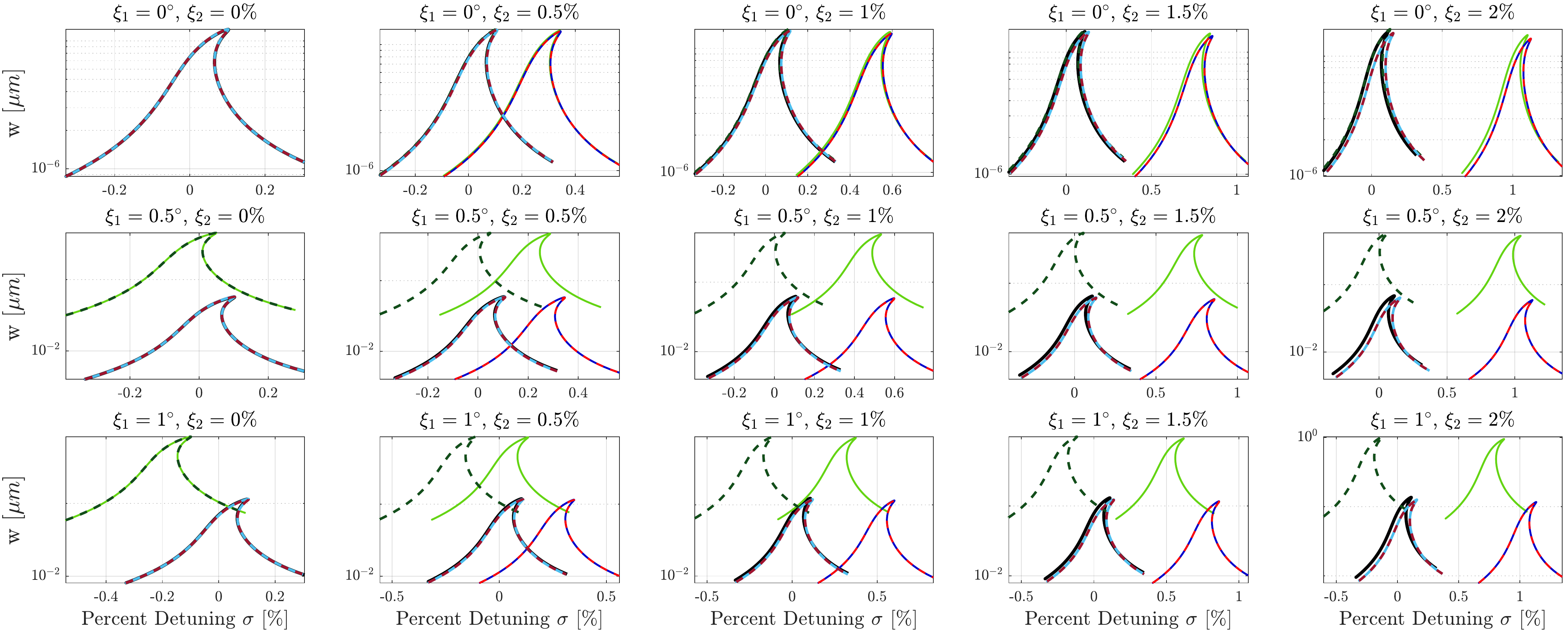} 
  \caption{Sense response (out of plane).}
  \label{subfig:figura_TS1_w}
\end{subfigure}
\caption{Frequency Responses and backbone curves for different defect amplitudes $\xi_1$ (shown in degrees) and $\xi_2$ (reported as percentage of the beam width) using the Harmonic Balance method (5 harmonics) for: ROM-d ({\ROMd}), DpROM-N0 ({\Nzero}), DpROM-N1t ({\Nunot}), DpROM-N1 ({\Nuno}), DpROM-N0-v ({\Nzerov}),  DpROM-N1t-v ({\Nunotv}) and  DpROM-N1-v ({\Nunov}). The displacements $u$ and $w$ of the center of the mass is shown (first harmonic coefficient of the Fourier series). For each plot, the percent detuning parameter $\sigma$ is referred to the corresponding FOM-d first eigenfrequency.}
\label{fig:figura_TS1_uw}
\end{figure}

The frequency response was obtained forcing the system in the center of the suspended mass with a nodal load directed along the x-direction, with amplitude $F=0.4 \, \mu$N, and using a Rayleigh damping matrix with $\alpha=105$ and $\beta=0$. Figure \ref{fig:figura_TS1_uw} reports the FRs around the first eigenfrequency of the system for the x-displacement $u$ (drive direction) and the z-displacement $w$ (sense direction) for all the combinations of $\xi_1=\{0^{\circ},\,0.5^{\circ},\,1^{\circ}\}$ and $\xi_2=\{0\%,\, 0.5\%,\, 1\%,\, 1.5\%,\, 2\% \}$.

\subsection{Tested models}
For the present study, we considered also a truncated version of DpROM-N1 (named DpROM-N1t) where we considered negligible in Eq. \eqref{eq:E2N} the strains of order $\mathcal{O}(\D_d\D^2)$. This further assumption allows us to drop all the terms multiplying $\Aiii$ or, equivalently, $\textbf{L}_3$, so that $\tenb{Q}{5d}{}$, $\tenb{Q}{5dd}{}$, $\tenb{Q}{6dd}{}$ and the last two terms in Eq. \eqref{subeq:Q4d} are null. Although introducing a new approximation, DpROM-N1t is significantly cheaper to construct and, as it will be shown, does not introduce any appreciable accuracy loss in our studies. For each of the presented pROMs then, we test the same models with the volume-integration correction described in Section \ref{sec:5.3}. We will address to these models appending the suffix ``-v'' to the name of the model itself (e.g. DpROM-N1t-v). To recap, the results for a total of 6 models are reported in the following: DpROM-N0(-v), DpROM-N1t(-v) and DpROM-N1(-v). Again, see Table \ref{tab:acronyms} for a quick reference.

\subsection{Results}
Considering first the effect of the wall angle defect only, it is apparent how DpROM-N0 performances quickly degrade as soon as the parameter $\xi_1$ is increased. This can be seen both in the error on the linear eigenfrequency and especially in the overestimated $w$-response, approximately one order of magnitude higher than the reference. This may be due to the fact that the S-shaped beams are specifically designed to minimize the cross-coupling between the drive (x-) and sense (z-) axes created by the wall angle, so that the $w$-response is so small (2 orders of magnitude lower than the $u$-response) that it cannot be accurately captured by DpROM-N0. The same observations can be made for DpROM-N0-v, as the wall angle defect by itself represent an isochoric transformation. The responses of all the other tested DpROMs show instead a perfect match with the reference when $\xi_2=0\%$.

If the tapering defect only is considered (i.e. with $\xi_1=0^{\circ}$), we observe that DpROM-N0/N1t/N1 have similar responses, with an error on the eigenfrequency that translates the whole response by an approximately constant $\Delta f$. The models with the volume-integration correction were then tested. If on the one hand DpROM-N0-v still presents relevant errors, on the other hand DpROM-N1t-v and DpROM-N1v show very accurate results in the full range of the tested $\xi_2$. Such an improvement was expected, as the volume changed by this defect affects the suspension beams dimensions, to which the eigenfrequencies of the system are very sensitive.

For the remaining cases, the trends observed for the parameters $\xi_1$ and $\xi_2$ individually mix. Notice that looking at some results (e.g. $u$-response for $\xi_1 = 1^{\circ}$, $\xi_2=0.5\%$), it may seem that DpROM-N0 gives  better results than DpROM-N1. This is however just a coincidence, as for DpROM-N0 the first defect shifts the first eigenfrequency to lower frequencies while the second defect to higher frequencies, so that the two errors in this case cancel out. Indeed, when the volume correction is used in DpROM-N0-v, only the first effect is observed, and the frequencies are shifted to the left.

\begin{figure}[t!]
  \centering
  \includegraphics[width=1\linewidth]{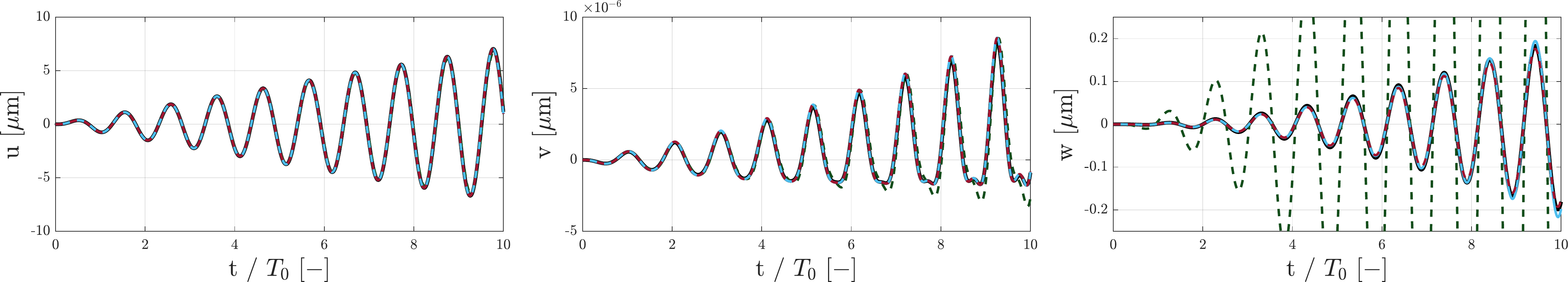}
  \caption{Transient response of the center of the suspended mass of the gyroscope for: ROM-d ({\ROMd}), DpROM-N0-v ({\Nzerov}),  DpROM-N1t-v ({\Nunotv}) and  DpROM-N1-v ({\Nunov}). The forcing is harmonic with period T$_0$. Case with $\xi_1 = 1^{\circ}$, $\xi_2=2\%$.}
  \label{fig:TS1_transient}
\end{figure}

In Fig. \ref{fig:TS1_transient} we also show the transient response of the forced node for ROM-d and DpROMs-v (case with $\xi_1 = 1^{\circ}$, $\xi_2=0.5\%$). Each model is forced at its own first resonance frequency $f_0$ (as it is usually the case for MEMS gyroscopes) with a harmonic forcing, taking 100 samples per period and for a time span equal to 10 times T$_0=1/f_0$, with $F=50\mu$N. The integration was carried out in Matlab with our in-house code, using a Newmark integration scheme. Looking at the responses along the three axes, we observe that the three DpROMs yield correct results but for DpROM-N0-v along the sense z-direction ($w$ component). Also, considering the z-response, we can see that DpROM-N1 is slightly better DpROM-N1t, fact that was not very visible in the FRs.

\subsection{Computational times}
Table \ref{tab:comptimes_TS1} reports the average time for the FR analyses and for the construction of the different models. To compare in terms of time ROM-d and the DpROMs, it is convenient to consider the \textit{variable costs} (T$_{var}$), i.e. the ones that have to be sustained for each new parameter realization, and the \textit{overhead costs} (T$_{oh}$), i.e. the ones sustained once and for all independently from the number of realizations. In the case of ROM-d we have that $\text{T}_{var}=\text{T}_{constr}+\text{T}_{sim}$, being $\text{T}_{constr}$ the time to construct the model (i.e. RB and tensors computation) and $\text{T}_{sim}$ the time for one simulation, while T$_{oh}=0$. For ROM-d indeed, there are no common overhead costs, but a new model must be constructed for each new realization of the parameters. In the case of DpROMs instead, we have that  $\text{T}_{var}^{p}=\text{T}_{sim}^p$ and $\text{T}_{oh}^p=\text{T}_{constr}^p$ (we use the superscript ``p'' to distinguish the parametric models from ROM-d). For the parametric models we have in fact to pay upfront the cost of model construction, which is generally more expensive than the one for ROM-d, but thereafter only the simulation cost must be sustained for each new case. The first trivial conclusion is then that there exist a number $\bar N$ of parameter realization above which DpROMs become convenient, that is:
\begin{equation}
N \geq \bar N = \bigg\lceil \frac{T_{oh}^p}{T_{var}-T_{var}^p} \bigg\rceil .
\end{equation}
For $\bar N$ to be positive and finite, it follows that
\begin{equation}
T_{var} > T_{var}^p \,\,\, \longleftrightarrow \,\,\, T_{constr} + T_{sim} > T_{sim}^p.
\label{eq:time_convenience}
\end{equation}
From Eq. \eqref{eq:time_convenience} it can be seen how the convenience of the parametric model over the non-parametric one depends on the relative weight between the simulation and construction times of the latter and the simulation time of the former, as it can be observed in Table \ref{tab:comptimes_TS1} looking at the different speedups\footnote{Speedups are computed considering the variable costs only, with respect to ROM-d.} for the FR and transient analyses.

That said, it is clearly difficult to draw general and definitive conclusions on the benefits of the two solutions, ROM-d and DpROMs, time-wise. In the experience of the authors, transient analysis offer the best gains, as simulation speed is very high, grows almost linearly with the simulation time span, and is less sensitive to the number of dofs than other kind of analysis, as the ones requiring continuation methods. When continuation is required, one could potentially find greater benefits in using a model with a low number of dofs, so that ROM-d could actually become the best choice. We remark however that for ROM-d we have to take into account also the construction cost as a variable cost, and that for large FE models the sole computation of structural eigenmodes can already take several minutes, making this cost very high.

\begin{table}[t]
\centering
\caption{Average computational times for the FR with the HB method and for the construction of each ROM (comprising the time to compute VMs, MDs, DSs and the tensors). ROM-d counts 9 dofs, while DpROMs 15 dofs. Apart from the construction costs, items for the DpROMs are clustered and averaged. Notice that the construction time for ROM-d is sustained for each new parameter realization and contributes to the \textit{variable} costs. DpROM names are reported by their respective suffix.}
\label{tab:comptimes_TS1}
\begin{tabular}{@{}lccccccc@{}}
\toprule
Model--II          	      & \textbf{ROM-d} & \textbf{N0} & \textbf{N1t} & \textbf{N1} & \textbf{N0-v} & \textbf{N1t-v} & \textbf{N1-v} \\ \midrule
\textbf{ROM construction} & 209 s          & 335 s       & 333 s        & 816 s       & 353 s         & 357 s          &  1,063 s \\ 
\textbf{HB FR}            &  21.7 s          &  \multicolumn{6}{c}{ 49.2 s }  \\
\textbf{Transient analysis}& 0.43 s         &  \multicolumn{6}{c}{0.51 s } \\
\textit{Overhead cost} & --          & 335 s       & 333 s        & 816 s       & 353 s         & 357 s          &  1,063 s \\
\textit{Variable cost (FR/transient)}&	230.7 s / 209.4 s         &  \multicolumn{6}{c}{ 49.2 s / 0.51 s}  \\
\textit{Speedup (FR/transient)}     &	-- / --            &  \multicolumn{6}{c}{$4.7\times$ / $410.6\times$} \\
\bottomrule
\end{tabular}
\end{table}

\section{Conclusions}
We presented a ROM for geometric nonlinearities that can parametrically describe a shape imperfection with respect to the nominal (blueprint) design, named for brevity DpROM. The imperfection is given by the superposition of user-defined defect shapes, whose amplitudes are parameters of the model and can be changed without reconstructing the model itself. This result has been made possible thanks to a polynomial representation of the internal forces resulting from a two-step deformation process (which brings the nominal geometry into the defected one and then into the deformed one) and from the approximation of the strains obtained by a Neumann expansion. The latter allowed to eliminate rational expressions under the hypothesis of small defects, so that the elastic internal forces are written as simple polynomials both with respect to the displacement field representing the defect and with respect to the actual displacement field. Using a Galerkin projection and a modal-based approach for selecting the RB, the reduced internal forces have been recast in tensorial form, where the linear, quadratic and cubic stiffness tensors are found to be functions of a parameter vector collecting the amplitudes of the defects imposed on the structure. Within this framework we tested different versions of the DpROM for different degrees of approximation. In particular, we have shown that the model we had previously developed using Budiansky's approach corresponds to the 0th-order expansion of our model, without volume integration correction (i.e. DpROM-N0). Finally, in the numerical studies we showed that the higher order approximation DpROM-N1 effectively leads to more accurate results and that for volume-changing defects a large improvement can be achieved by approximating the tensor integral over the real volume of the defective geometry (DpROM-N1-v). The truncated version DpROM-N1t(-v) was also presented, which has almost the same accuracy as its complete counterpart, but without the need to construct tensors with dimensionality higher than four. The computational costs were then critically discussed, taking into account different types of analysis. In particular, we showed that in transient studies we can usually expect very high speedups from the parametric models. In the case of FR analysis, which we used to assess the quality of the solutions over a range of frequencies as an alternative to multiple time analyses, the gains will be more contained. In this context, to reduce the dofs of both the parametric and non-parametric ROMs and make FR analysis faster and thus closer to transient analysis in terms of time and speedups, we think that an a priori selection of the RB vectors and hyperreduction strategies would actually be very beneficial, and they can actually constitute the spur for future investigation.

\section*{Acknowledgements}
The authors want to express their gratitude to the AMS division at STMicroelectronics in Cornaredo (MI, Italy), for their support in this work.

\newpage
\appendix

\section{$\textbf{L}_1$, $\textbf{L}_2$ and $\textbf{L}_3$ matrices}
\label{appendixA}
We report in Tables \ref{tab:L2D} and \ref{tab:L3D} the expressions for the matrices $\textbf{L}_1$, $\textbf{L}_2$ and $\textbf{L}_3$ defined in Eqs. \ref{eq:Ls}.
\begin{table}[h!]
\centering
\caption{Elements $L^{(1)}_{ijk}$, $L^{(2)}_{ijk}$ and $L^{(3)}_{ijkl}$ of the sparse $3 \times 4 \times 4$  matrices $\textbf{L}_1$, $\textbf{L}_2$ and of the sparse $3 \times 4 \times 4 \times 4$  matrix $\textbf{L}_3$, respectively, in the 2D case (name subscripts are moved to superscripts to avoid confusion with the indexes).}
\label{tab:L2D}
\resizebox{\textwidth}{!}{%
\begin{tabular}{llllllll}
\hline
$L^{(1)}_{111}=1$, & $L^{(1)}_{321}=1$, & $L^{(1)}_{312}=1$, & $L^{(1)}_{222}=1$, & $L^{(1)}_{123}=1$, & $L^{(1)}_{343}=1$, & $L^{(1)}_{324}=1$, & $L^{(1)}_{244}=1$. \\ \hline
$L^{(2)}_{111}=1$, & $L^{(2)}_{331}=1$, & $L^{(2)}_{312}=1$, & $L^{(2)}_{232}=1$, & $L^{(2)}_{123}=1$, & $L^{(2)}_{343}=1$, & $L^{(2)}_{324}=1$, & $L^{(2)}_{244}=1$. \\ \hline
$L^{(3)}_{1111}= 1$, & $L^{(3)}_{3211}= \frac{1}{2}$, & $L^{(3)}_{3121}= \frac{1}{2}$, & $L^{(3)}_{1331}= 1$, & $L^{(3)}_{3431}= \frac{1}{2}$, & $L^{(3)}_{3341}= \frac{1}{2}$, & $L^{(3)}_{3112}= 1$, & $L^{(3)}_{2212}= \frac{1}{2}$, \\
$L^{(3)}_{2122}= \frac{1}{2}$, & $L^{(3)}_{3332}= 1$, & $L^{(3)}_{2432}= \frac{1}{2}$, & $L^{(3)}_{2342}= \frac{1}{2}$, & $L^{(3)}_{1213}= \frac{1}{2}$, & $L^{(3)}_{1123}= \frac{1}{2}$, & $L^{(3)}_{3223}= 1$, & $L^{(3)}_{1433}= \frac{1}{2}$, \\
$L^{(3)}_{1343}= \frac{1}{2}$, & $L^{(3)}_{3443}= 1$, & $L^{(3)}_{3214}= \frac{1}{2}$, & $L^{(3)}_{3124}= \frac{1}{2}$, & $L^{(3)}_{2224}= 1$, & $L^{(3)}_{3434}= \frac{1}{2}$, & $L^{(3)}_{3344}= \frac{1}{2}$, & $L^{(3)}_{2444}= 1$. \\ \hline
\end{tabular}%
}
\end{table}
\begin{table}[h!]
\centering
\caption{Elements $L^{(1)}_{ijk}$, $L^{(2)}_{ijk}$ and $L^{(3)}_{ijkl}$ of the sparse $6 \times 9 \times 9$  matrices $\textbf{L}_1$, $\textbf{L}_2$ and of the sparse $6 \times 9 \times 9 \times 9$  matrix $\textbf{L}_3$, respectively, in the 3D case (name subscripts are moved to superscripts to avoid confusion with the indexes).}
\label{tab:L3D}
\resizebox{\textwidth}{!}{%
\begin{tabular}{lllllllll}
\hline
$L^{(1)}_{111}=1$, & $L^{(1)}_{421}=1$, & $L^{(1)}_{531}=1$, & $L^{(1)}_{412}=1$, & $L^{(1)}_{222}=1$, & $L^{(1)}_{632}=1$, & $L^{(1)}_{513}=1$, & $L^{(1)}_{623}=1$, & $L^{(1)}_{333}=1$, \\
$L^{(1)}_{144}=1$, & $L^{(1)}_{454}=1$, & $L^{(1)}_{564}=1$, & $L^{(1)}_{445}=1$, & $L^{(1)}_{255}=1$, & $L^{(1)}_{665}=1$, & $L^{(1)}_{546}=1$, & $L^{(1)}_{656}=1$, & $L^{(1)}_{366}=1$, \\
$L^{(1)}_{177}=1$, & $L^{(1)}_{487}=1$, & $L^{(1)}_{597}=1$, & $L^{(1)}_{478}=1$, & $L^{(1)}_{288}=1$, & $L^{(1)}_{698}=1$, & $L^{(1)}_{579}=1$, & $L^{(1)}_{689}=1$, & $L^{(1)}_{399}=1$. \\ \hline
$L^{(2)}_{111}=1$, & $L^{(2)}_{441}=1$, & $L^{(2)}_{571}=1$, & $L^{(2)}_{412}=1$, & $L^{(2)}_{242}=1$, & $L^{(2)}_{672}=1$, & $L^{(2)}_{513}=1$, & $L^{(2)}_{643}=1$, & $L^{(2)}_{373}=1$, \\
$L^{(2)}_{124}=1$, & $L^{(2)}_{454}=1$, & $L^{(2)}_{584}=1$, & $L^{(2)}_{425}=1$, & $L^{(2)}_{255}=1$, & $L^{(2)}_{685}=1$, & $L^{(2)}_{526}=1$, & $L^{(2)}_{656}=1$, & $L^{(2)}_{386}=1$, \\
$L^{(2)}_{137}=1$, & $L^{(2)}_{467}=1$, & $L^{(2)}_{597}=1$, & $L^{(2)}_{438}=1$, & $L^{(2)}_{268}=1$, & $L^{(2)}_{698}=1$, & $L^{(2)}_{539}=1$, & $L^{(2)}_{669}=1$, & $L^{(2)}_{399}=1$. \\ \hline
$L^{(3)}_{1111}=1$, & $L^{(3)}_{4211}=\frac{1}{2}$, & $L^{(3)}_{5311}=\frac{1}{2}$, & $L^{(3)}_{4121}=\frac{1}{2}$, & $L^{(3)}_{5131}=\frac{1}{2}$, & $L^{(3)}_{1441}=1$, & $L^{(3)}_{4541}=\frac{1}{2}$, & $L^{(3)}_{5641}=\frac{1}{2}$, & $L^{(3)}_{4451}=\frac{1}{2}$, \\
$L^{(3)}_{5461}=\frac{1}{2}$, & $L^{(3)}_{1771}=1$, & $L^{(3)}_{4871}=\frac{1}{2}$, & $L^{(3)}_{5971}=\frac{1}{2}$, & $L^{(3)}_{4781}=\frac{1}{2}$, & $L^{(3)}_{5791}=\frac{1}{2}$, & $L^{(3)}_{4112}=1$, & $L^{(3)}_{2212}=\frac{1}{2}$, & $L^{(3)}_{6312}=\frac{1}{2}$, \\
$L^{(3)}_{2122}=\frac{1}{2}$, & $L^{(3)}_{6132}=\frac{1}{2}$, & $L^{(3)}_{4442}=1$, & $L^{(3)}_{2542}=\frac{1}{2}$, & $L^{(3)}_{6642}=\frac{1}{2}$, & $L^{(3)}_{2452}=\frac{1}{2}$, & $L^{(3)}_{6462}=\frac{1}{2}$, & $L^{(3)}_{4772}=1$, & $L^{(3)}_{2872}=\frac{1}{2}$, \\
$L^{(3)}_{6972}=\frac{1}{2}$, & $L^{(3)}_{2782}=\frac{1}{2}$, & $L^{(3)}_{6792}=\frac{1}{2}$, & $L^{(3)}_{5113}=1$, & $L^{(3)}_{6213}=\frac{1}{2}$, & $L^{(3)}_{3313}=\frac{1}{2}$, & $L^{(3)}_{6123}=\frac{1}{2}$, & $L^{(3)}_{3133}=\frac{1}{2}$, & $L^{(3)}_{5443}=1$, \\
$L^{(3)}_{6543}=\frac{1}{2}$, & $L^{(3)}_{3643}=\frac{1}{2}$, & $L^{(3)}_{6453}=\frac{1}{2}$, & $L^{(3)}_{3463}=\frac{1}{2}$, & $L^{(3)}_{5773}=1$, & $L^{(3)}_{6873}=\frac{1}{2}$, & $L^{(3)}_{3973}=\frac{1}{2}$, & $L^{(3)}_{6783}=\frac{1}{2}$, & $L^{(3)}_{3793}=\frac{1}{2}$, \\
$L^{(3)}_{1214}=\frac{1}{2}$, & $L^{(3)}_{1124}=\frac{1}{2}$, & $L^{(3)}_{4224}=1$, & $L^{(3)}_{5324}=\frac{1}{2}$, & $L^{(3)}_{5234}=\frac{1}{2}$, & $L^{(3)}_{1544}=\frac{1}{2}$, & $L^{(3)}_{1454}=\frac{1}{2}$, & $L^{(3)}_{4554}=1$, & $L^{(3)}_{5654}=\frac{1}{2}$, \\
$L^{(3)}_{5564}=\frac{1}{2}$, & $L^{(3)}_{1874}=\frac{1}{2}$, & $L^{(3)}_{1784}=\frac{1}{2}$, & $L^{(3)}_{4884}=1$, & $L^{(3)}_{5984}=\frac{1}{2}$, & $L^{(3)}_{5894}=\frac{1}{2}$, & $L^{(3)}_{4215}=\frac{1}{2}$, & $L^{(3)}_{4125}=\frac{1}{2}$, & $L^{(3)}_{2225}=1$, \\
$L^{(3)}_{6325}=\frac{1}{2}$, & $L^{(3)}_{6235}=\frac{1}{2}$, & $L^{(3)}_{4545}=\frac{1}{2}$, & $L^{(3)}_{4455}=\frac{1}{2}$, & $L^{(3)}_{2555}=1$, & $L^{(3)}_{6655}=\frac{1}{2}$, & $L^{(3)}_{6565}=\frac{1}{2}$, & $L^{(3)}_{4875}=\frac{1}{2}$, & $L^{(3)}_{4785}=\frac{1}{2}$, \\
$L^{(3)}_{2885}=1$, & $L^{(3)}_{6985}=\frac{1}{2}$, & $L^{(3)}_{6895}=\frac{1}{2}$, & $L^{(3)}_{5216}=\frac{1}{2}$, & $L^{(3)}_{5126}=\frac{1}{2}$, & $L^{(3)}_{6226}=1$, & $L^{(3)}_{3326}=\frac{1}{2}$, & $L^{(3)}_{3236}=\frac{1}{2}$, & $L^{(3)}_{5546}=\frac{1}{2}$, \\
$L^{(3)}_{5456}=\frac{1}{2}$, & $L^{(3)}_{6556}=1$, & $L^{(3)}_{3656}=\frac{1}{2}$, & $L^{(3)}_{3566}=\frac{1}{2}$, & $L^{(3)}_{5876}=\frac{1}{2}$, & $L^{(3)}_{5786}=\frac{1}{2}$, & $L^{(3)}_{6886}=1$, & $L^{(3)}_{3986}=\frac{1}{2}$, & $L^{(3)}_{3896}=\frac{1}{2}$, \\
$L^{(3)}_{1317}=\frac{1}{2}$, & $L^{(3)}_{4327}=\frac{1}{2}$, & $L^{(3)}_{1137}=\frac{1}{2}$, & $L^{(3)}_{4237}=\frac{1}{2}$, & $L^{(3)}_{5337}=1$, & $L^{(3)}_{1647}=\frac{1}{2}$, & $L^{(3)}_{4657}=\frac{1}{2}$, & $L^{(3)}_{1467}=\frac{1}{2}$, & $L^{(3)}_{4567}=\frac{1}{2}$, \\
$L^{(3)}_{5667}=1$, & $L^{(3)}_{1977}=\frac{1}{2}$, & $L^{(3)}_{4987}=\frac{1}{2}$, & $L^{(3)}_{1797}=\frac{1}{2}$, & $L^{(3)}_{4897}=\frac{1}{2}$, & $L^{(3)}_{5997}=1$, & $L^{(3)}_{4318}=\frac{1}{2}$, & $L^{(3)}_{2328}=\frac{1}{2}$, & $L^{(3)}_{4138}=\frac{1}{2}$, \\
$L^{(3)}_{2238}=\frac{1}{2}$, & $L^{(3)}_{6338}=1$, & $L^{(3)}_{4648}=\frac{1}{2}$, & $L^{(3)}_{2658}=\frac{1}{2}$, & $L^{(3)}_{4468}=\frac{1}{2}$, & $L^{(3)}_{2568}=\frac{1}{2}$, & $L^{(3)}_{6668}=1$, & $L^{(3)}_{4978}=\frac{1}{2}$, & $L^{(3)}_{2988}=\frac{1}{2}$, \\
$L^{(3)}_{4798}=\frac{1}{2}$, & $L^{(3)}_{2898}=\frac{1}{2}$, & $L^{(3)}_{6998}=1$, & $L^{(3)}_{5319}=\frac{1}{2}$, & $L^{(3)}_{6329}=\frac{1}{2}$, & $L^{(3)}_{5139}=\frac{1}{2}$, & $L^{(3)}_{6239}=\frac{1}{2}$, & $L^{(3)}_{3339}=1$, & $L^{(3)}_{5649}=\frac{1}{2}$, \\
$L^{(3)}_{6659}=\frac{1}{2}$, & $L^{(3)}_{5469}=\frac{1}{2}$, & $L^{(3)}_{6569}=\frac{1}{2}$, & $L^{(3)}_{3669}=1$, & $L^{(3)}_{5979}=\frac{1}{2}$, & $L^{(3)}_{6989}=\frac{1}{2}$, & $L^{(3)}_{5799}=\frac{1}{2}$, & $L^{(3)}_{6899}=\frac{1}{2}$, & $L^{(3)}_{3999}=1$. \\ \hline
\end{tabular}%
}
\end{table}

\section{Tangent stiffness matrix derivatives}
\label{appendixB}
The virtual variation wrt $\ub$ of the internal elastic forces as defined in Eq. \eqref{eq:fint_full} writes
\begin{equation}
\begin{split}
\delta \f_{int} &= \int_{V_o} \left[ \G^T \left( \Hb + \A + \Aii + 2\Aiii\A\right)^T \Cb \left( \Hb + \frac 12 \A + \Aii + \Aiii\A \right) \G \delta \ub + \right. \\
&+ \G^T \left( \Hb + \A + \Aii + 2\Aiii\A\right)^T \Cb \left( \frac 12 \delta\A + \Aiii\delta\A \right) \G \ub + \\
&+ \left. \G^T \left( \delta\A + 2\Aiii\delta\A\right)^T \Cb \left( \Hb + \frac 12 \A + \Aii + \Aiii\A \right) \G \ub \right] \text{ d}V_o.
\end{split}
\end{equation}
Recalling that $\A\delta\thb=\delta\A\thb$, we can write
\begin{equation}
\begin{split}
\delta \f_{int} &= \int_{V_o} \left[ \G^T \left( \Hb + \A + \Aii + 2\Aiii\A\right)^T \Cb \left( \Hb + \A + \Aii + 2\Aiii\A \right) \G \delta \ub + \right. \\
&+ \G^T \delta\A^T \underbrace{ \left( \I + 2\Aiii^T \right) \Cb \left( \Hb + \frac 12 \A + \Aii + \Aiii\A \right) \G \ub }_{ \textbf{N}(\ub,\ub_d) } \Big] \text{ d}V_o,
\end{split}
\end{equation}
where the second term on the right-hand side can be rewritten to put in evidence the displacement virtual variation $\delta \ub$ as
\begin{equation}
\delta f_{I}^{\prime\prime} = \int_{V_o} G_{iI} L_{1jik} G_{kl} \delta u_l N_j \text{ d}V_o,
\end{equation}
where Einstein notation was used for convenience. The tangent stiffness matrix therefore writes:
\begin{equation}
\textbf{K}_t = \textbf{K} ^{\prime} + \textbf{K} ^{\prime\prime}
\label{eq:tan_stiff_mat}
\end{equation}
where
\begin{equation}
\textbf{K} ^{\prime} = \int_{V_o} \G^T \left( \Hb + \A + \Aii + 2\Aiii\A\right)^T \Cb \left( \Hb + \A + \Aii + 2\Aiii\A \right) \G\text{ d}V_o ,
\end{equation}
\begin{equation}
K_{IJ}^{\prime\prime} = \int_{V_o} G_{iI} L_{1jik} G_{kJ} N_j \text{ d}V_o.
\end{equation}
Substituting $\ub=\Phib_i\eta_i$ and $\ub_d = \Vd_j\xi_j$ in Eq. \eqref{eq:tan_stiff_mat}, taking the derivative wrt either $\eta_i$ and/or $\xi_j$ and evaluating the resulting expressions at equilibrium and with zero defect amplitudes, as required by Eqs. \eqref{eq:MD}--\eqref{eq:DS2}, we can write the derivatives of $\textbf{K}_t$ as:
\begin{subequations}
\begin{align}
    \frac{\partial \textbf{K}_t}{\partial \eta_j} \bigg|_0 &= \int_{V_o} \left[  \G^T \left( \Hb^T\Cb{\A}_j + {\A}_j^T\Cb\Hb  \right) \G + \G\cdot_{11} \left[\left(\LL \cdot \G \right) \cdot_{11} \left(\Cb \Hb \G \Phib_i \right) \right] \right] \text{ d}V_o,\\
    \frac{\partial \textbf{K}_t}{\partial \xi_j} \bigg|_0 &= \int_{V_o} \G^T \left( \Hb^T\Cb{\Aii}_j + {\Aii}_j^T\Cb\Hb \right) \G \text{ d}V_o,\\
    \begin{split}
    \frac{\partial^2 \textbf{K}_t}{\partial \eta_j \partial \xi_k} \bigg|_0 &= \int_{V_o} \left[ \G^T \left( {\Aii}_k^T\Cb{\A}_j + {\A}_j^T\Cb{\Aii}_k + 2{\A}_j^T{\Aiii}_k^T\Cb\Hb + 2\Hb^T\Cb{\Aiii}_k{\A}_j \right) \G + \right.\\
    &+ \G\cdot_{11} \left[\left(\LL \cdot \G \right) \cdot_{11} \left(\left(\Cb \Aii + 2\Aiii^T\Cb\Hb\right) \G \Phib_i \right) \right] \Big] \text{ d}V_o ,
    \end{split} \\
    \frac{\partial^2 \textbf{K}_t}{\partial \xi_j \partial \xi_k} \bigg|_0 &= \int_{V_o} \G^T \left( {\Aii}_j^T\Cb{\Aii}_k + {\Aii}_k^T\Cb{\Aii}_j \right) \G \text{ d}V_o,
\end{align}
\end{subequations}
where, recalling that $\eta_i$ and $\xi_j$ are scalars, we used $\A(\G\Phib_i\eta_i) = {\A}_i \eta_i$ and $\Aii(\G\Vd_j\xi_j) = {\Aii}_j \xi_j$ (same for $\Aiii$) to avoid a cumbersome notation, and where $\cdot_{ij}$ denotes the contraction of the i-th dimension of the first term with the j-th dimension of the second term.

\bibliography{mybibfile}

\begin{thebibliography}{60}
\expandafter\ifx\csname natexlab\endcsname\relax\def\natexlab#1{#1}\fi
\providecommand{\url}[1]{\texttt{#1}}
\providecommand{\href}[2]{#2}
\providecommand{\path}[1]{#1}
\providecommand{\DOIprefix}{doi:}
\providecommand{\ArXivprefix}{arXiv:}
\providecommand{\URLprefix}{URL: }
\providecommand{\Pubmedprefix}{pmid:}
\providecommand{\doi}[1]{\href{http://dx.doi.org/#1}{\path{#1}}}
\providecommand{\Pubmed}[1]{\href{pmid:#1}{\path{#1}}}
\providecommand{\bibinfo}[2]{#2}
\ifx\xfnm\relax \def\xfnm[#1]{\unskip,\space#1}\fi
\bibitem[{Marconi et~al.(2020)Marconi, Tiso, and Braghin}]{Marconi2020}
\bibinfo{author}{J.~Marconi}, \bibinfo{author}{P.~Tiso},
  \bibinfo{author}{F.~Braghin},
\newblock \bibinfo{title}{{A nonlinear reduced order model with parametrized
  shape defects}},
\newblock \bibinfo{journal}{Computer Methods in Applied Mechanics and
  Engineering} \bibinfo{volume}{360} (\bibinfo{year}{2020})
  \bibinfo{pages}{112785}.
\bibitem[{Belytschko et~al.(2013)Belytschko, Liu, Moran, and
  Elkhodary}]{Belytschko2014}
\bibinfo{author}{T.~Belytschko}, \bibinfo{author}{W.~K. Liu},
  \bibinfo{author}{B.~Moran}, \bibinfo{author}{K.~Elkhodary},
  \bibinfo{title}{Nonlinear finite elements for continua and structures},
  \bibinfo{publisher}{John wiley \& sons}, \bibinfo{year}{2013}.
\bibitem[{Toselli and Widlund(2005)}]{Toselli2005}
\bibinfo{author}{A.~Toselli}, \bibinfo{author}{O.~B. Widlund},
  \bibinfo{title}{{Domain Decomposition Methods — Algorithms and Theory}},
  volume~\bibinfo{volume}{34} of \textit{\bibinfo{series}{Springer Series in
  Computational Mathematics}}, \bibinfo{publisher}{Springer Berlin Heidelberg},
  \bibinfo{address}{Berlin, Heidelberg}, \bibinfo{year}{2005}.
  \DOIprefix\doi{10.1007/b137868}.
\bibitem[{Klerk et~al.(2008)Klerk, Rixen, and Voormeeren}]{Klerk2008}
\bibinfo{author}{D.~D. Klerk}, \bibinfo{author}{D.~J. Rixen},
  \bibinfo{author}{S.~N. Voormeeren},
\newblock \bibinfo{title}{{General Framework for Dynamic Substructuring:
  History, Review and Classification of Techniques}},
\newblock \bibinfo{journal}{AIAA Journal} \bibinfo{volume}{46}
  (\bibinfo{year}{2008}) \bibinfo{pages}{1169--1181}.
\bibitem[{Farhat and Roux(1991)}]{Farhat1991}
\bibinfo{author}{C.~Farhat}, \bibinfo{author}{F.-X. Roux},
\newblock \bibinfo{title}{{A method of finite element tearing and
  interconnecting and its parallel solution algorithm}},
\newblock \bibinfo{journal}{International Journal for Numerical Methods in
  Engineering} \bibinfo{volume}{32} (\bibinfo{year}{1991})
  \bibinfo{pages}{1205--1227}.
\bibitem[{Guyan(1965)}]{GUYAN1965}
\bibinfo{author}{R.~J. Guyan},
\newblock \bibinfo{title}{{Reduction of stiffness and mass matrices}},
\newblock \bibinfo{journal}{AIAA Journal} \bibinfo{volume}{3}
  (\bibinfo{year}{1965}) \bibinfo{pages}{380--380}.
\bibitem[{He and Fu(2001)}]{He2001}
\bibinfo{author}{J.~He}, \bibinfo{author}{Z.-F. Fu}, \bibinfo{title}{{Modal
  Analysis}}, \bibinfo{publisher}{Elsevier}, \bibinfo{year}{2001}.
  \DOIprefix\doi{10.1016/B978-0-7506-5079-3.X5000-1}.
\bibitem[{Craig and Bampton(1968)}]{Craig1968}
\bibinfo{author}{R.~Craig}, \bibinfo{author}{M.~Bampton},
\newblock \bibinfo{title}{{Coupling of Substructures for Dynamic Analyses}}
  \bibinfo{volume}{6} (\bibinfo{year}{1968}) \bibinfo{pages}{1313--1319}.
\bibitem[{Rubin(1975)}]{Rubin1975}
\bibinfo{author}{S.~Rubin},
\newblock \bibinfo{title}{{Improved Component-Mode Representation for
  Structural Dynamic Analysis}},
\newblock \bibinfo{journal}{AIAA Journal} \bibinfo{volume}{13}
  (\bibinfo{year}{1975}) \bibinfo{pages}{995--1006}.
\bibitem[{Pichler et~al.(2017)Pichler, Witteveen, and Fischer}]{Pichler2019}
\bibinfo{author}{F.~Pichler}, \bibinfo{author}{W.~Witteveen},
  \bibinfo{author}{P.~Fischer},
\newblock \bibinfo{title}{{Reduced-Order Modeling of Preloaded Bolted
  Structures in Multibody Systems by the Use of Trial Vector Derivatives}},
\newblock \bibinfo{journal}{Journal of Computational and Nonlinear Dynamics}
  \bibinfo{volume}{12} (\bibinfo{year}{2017}) \bibinfo{pages}{051032}.
\bibitem[{Blockmans et~al.(2015)Blockmans, Tamarozzi, Naets, and
  Desmet}]{Blockmans2015}
\bibinfo{author}{B.~Blockmans}, \bibinfo{author}{T.~Tamarozzi},
  \bibinfo{author}{F.~Naets}, \bibinfo{author}{W.~Desmet},
\newblock \bibinfo{title}{{A nonlinear parametric model reduction method for
  efficient gear contact simulations}},
\newblock \bibinfo{journal}{International Journal for Numerical Methods in
  Engineering} \bibinfo{volume}{102} (\bibinfo{year}{2015})
  \bibinfo{pages}{1162--1191}.
\bibitem[{Balajewicz et~al.(2015)Balajewicz, Amsallem, and
  Farhat}]{Balajewicz2015}
\bibinfo{author}{M.~Balajewicz}, \bibinfo{author}{D.~Amsallem},
  \bibinfo{author}{C.~Farhat},
\newblock \bibinfo{title}{{Projection-based model reduction for contact
  problems}},
\newblock \bibinfo{journal}{International Journal for Numerical Methods in
  Engineering} \bibinfo{volume}{106} (\bibinfo{year}{2015})
  \bibinfo{pages}{644--663}.
\bibitem[{G{\'{e}}radin and Rixen(2016)}]{Geradin2016}
\bibinfo{author}{M.~G{\'{e}}radin}, \bibinfo{author}{D.~J. Rixen},
\newblock \bibinfo{title}{{A ‘nodeless' dual superelement formulation for
  structural and multibody dynamics application to reduction of contact
  problems}},
\newblock \bibinfo{journal}{International Journal for Numerical Methods in
  Engineering} \bibinfo{volume}{106} (\bibinfo{year}{2016})
  \bibinfo{pages}{773--798}.
\bibitem[{{Mehrdad Pourkiaee} and Zucca(2019)}]{MehrdadPourkiaee2019}
\bibinfo{author}{S.~{Mehrdad Pourkiaee}}, \bibinfo{author}{S.~Zucca},
\newblock \bibinfo{title}{{A Reduced Order Model for Nonlinear Dynamics of
  Mistuned Bladed Disks with Shroud Friction Contacts}},
\newblock \bibinfo{journal}{Journal of Engineering for Gas Turbines and Power}
  \bibinfo{volume}{141} (\bibinfo{year}{2019}) \bibinfo{pages}{1--13}.
\bibitem[{Ghavamian et~al.(2017)Ghavamian, Tiso, and Simone}]{Ghavamian2017}
\bibinfo{author}{F.~Ghavamian}, \bibinfo{author}{P.~Tiso},
  \bibinfo{author}{A.~Simone},
\newblock \bibinfo{title}{{POD–DEIM model order reduction for strain
  softening viscoplasticity}},
\newblock \bibinfo{journal}{Computer Methods in Applied Mechanics and
  Engineering} \bibinfo{volume}{317} (\bibinfo{year}{2017})
  \bibinfo{pages}{458--479}.
\bibitem[{Wu et~al.(2019)Wu, Tiso, Tatsis, Chatzi, and van Keulen}]{Wu2019}
\bibinfo{author}{L.~Wu}, \bibinfo{author}{P.~Tiso},
  \bibinfo{author}{K.~Tatsis}, \bibinfo{author}{E.~Chatzi},
  \bibinfo{author}{F.~van Keulen},
\newblock \bibinfo{title}{{A modal derivatives enhanced Rubin substructuring
  method for geometrically nonlinear multibody systems}},
\newblock \bibinfo{journal}{Multibody System Dynamics} \bibinfo{volume}{45}
  (\bibinfo{year}{2019}) \bibinfo{pages}{57--85}.
\bibitem[{Wu et~al.(2018)Wu, Tiso, and van Keulen}]{Wu2018}
\bibinfo{author}{L.~Wu}, \bibinfo{author}{P.~Tiso}, \bibinfo{author}{F.~van
  Keulen},
\newblock \bibinfo{title}{{Interface Reduction with Multilevel Craig–Bampton
  Substructuring for Component Mode Synthesis}},
\newblock \bibinfo{journal}{AIAA Journal}  (\bibinfo{year}{2018})
  \bibinfo{pages}{1--15}.
\bibitem[{Vizzaccaro et~al.(2020)Vizzaccaro, Salles, and
  Touz{\'{e}}}]{Vizzaccaro2020}
\bibinfo{author}{A.~Vizzaccaro}, \bibinfo{author}{L.~Salles},
  \bibinfo{author}{C.~Touz{\'{e}}},
\newblock \bibinfo{title}{{Comparison of nonlinear mappings for reduced-order
  modelling of vibrating structures: normal form theory and quadratic manifold
  method with modal derivatives}},
\newblock \bibinfo{journal}{Nonlinear Dynamics}  (\bibinfo{year}{2020}).
\bibitem[{Jain et~al.(2018)Jain, Tiso, and Haller}]{Jain2018ssm}
\bibinfo{author}{S.~Jain}, \bibinfo{author}{P.~Tiso},
  \bibinfo{author}{G.~Haller},
\newblock \bibinfo{title}{{Exact nonlinear model reduction for a von
  K{\'{a}}rm{\'{a}}n beam: Slow-fast decomposition and spectral submanifolds}},
\newblock \bibinfo{journal}{Journal of Sound and Vibration}
  \bibinfo{volume}{423} (\bibinfo{year}{2018}) \bibinfo{pages}{195--211}.
\bibitem[{Ponsioen et~al.(2020)Ponsioen, Jain, and Haller}]{Ponsioen2020}
\bibinfo{author}{S.~Ponsioen}, \bibinfo{author}{S.~Jain},
  \bibinfo{author}{G.~Haller},
\newblock \bibinfo{title}{{Model reduction to spectral submanifolds and
  forced-response calculation in high-dimensional mechanical systems}},
\newblock \bibinfo{journal}{Journal of Sound and Vibration}
  \bibinfo{volume}{488} (\bibinfo{year}{2020}) \bibinfo{pages}{115640}.
\bibitem[{Perez et~al.(2017)Perez, Bartram, Beberniss, Wiebe, and
  Spottswood}]{Perez2017}
\bibinfo{author}{R.~Perez}, \bibinfo{author}{G.~Bartram},
  \bibinfo{author}{T.~Beberniss}, \bibinfo{author}{R.~Wiebe},
  \bibinfo{author}{S.~M. Spottswood},
\newblock \bibinfo{title}{{Calibration of aero-structural reduced order models
  using full-field experimental measurements}},
\newblock \bibinfo{journal}{Mechanical Systems and Signal Processing}
  \bibinfo{volume}{86} (\bibinfo{year}{2017}) \bibinfo{pages}{49--65}.
\bibitem[{Touz{\'{e}} et~al.(2014)Touz{\'{e}}, Vidrascu, and
  Chapelle}]{Touze2014}
\bibinfo{author}{C.~Touz{\'{e}}}, \bibinfo{author}{M.~Vidrascu},
  \bibinfo{author}{D.~Chapelle},
\newblock \bibinfo{title}{{Direct finite element computation of non-linear
  modal coupling coefficients for reduced-order shell models}},
\newblock \bibinfo{journal}{Computational Mechanics} \bibinfo{volume}{54}
  (\bibinfo{year}{2014}) \bibinfo{pages}{567--580}.
\bibitem[{Hollkamp and Gordon(2008)}]{Hollkamp2008}
\bibinfo{author}{J.~J. Hollkamp}, \bibinfo{author}{R.~W. Gordon},
\newblock \bibinfo{title}{{Reduced-order models for nonlinear response
  prediction: Implicit condensation and expansion}},
\newblock \bibinfo{journal}{Journal of Sound and Vibration}
  \bibinfo{volume}{318} (\bibinfo{year}{2008}) \bibinfo{pages}{1139--1153}.
\bibitem[{Kuether et~al.(2015)Kuether, Deaner, Hollkamp, and
  Allen}]{Kuether2015}
\bibinfo{author}{R.~J. Kuether}, \bibinfo{author}{B.~J. Deaner},
  \bibinfo{author}{J.~J. Hollkamp}, \bibinfo{author}{M.~S. Allen},
\newblock \bibinfo{title}{{Evaluation of geometrically nonlinear reduced-order
  models with nonlinear normal modes}},
\newblock \bibinfo{journal}{AIAA Journal} \bibinfo{volume}{52}
  (\bibinfo{year}{2015}) \bibinfo{pages}{3273--3285}.
\bibitem[{Amabili(2013)}]{Amabili2013}
\bibinfo{author}{M.~Amabili},
\newblock \bibinfo{title}{{Reduced-order models for nonlinear vibrations, based
  on natural modes: the case of the circular cylindrical shell}},
\newblock \bibinfo{journal}{Philosophical Transactions of the Royal Society A:
  Mathematical, Physical and Engineering Sciences} \bibinfo{volume}{371}
  (\bibinfo{year}{2013}) \bibinfo{pages}{20120474}.
\bibitem[{Mignolet et~al.(2013)Mignolet, Przekop, Rizzi, and
  Spottswood}]{Mignolet2013}
\bibinfo{author}{M.~P. Mignolet}, \bibinfo{author}{A.~Przekop},
  \bibinfo{author}{S.~A. Rizzi}, \bibinfo{author}{S.~M. Spottswood},
\newblock \bibinfo{title}{{A review of indirect/non-intrusive reduced order
  modeling of nonlinear geometric structures}},
\newblock \bibinfo{journal}{Journal of Sound and Vibration}
  \bibinfo{volume}{332} (\bibinfo{year}{2013}) \bibinfo{pages}{2437--2460}.
\bibitem[{Jain(2015)}]{Jain2015}
\bibinfo{author}{S.~Jain}, \bibinfo{title}{{Model Order Reduction for
  Non-linear Structural Dynamics}}, \bibinfo{year}{2015}. \URLprefix
  \url{http://resolver.tudelft.nl/uuid:cb1d7058-2cfa-439a-bb2f-22a6b0e5bb2a}.
\bibitem[{Lu et~al.(2019)Lu, Jin, Chen, Yang, Hou, Zhang, Li, and Fu}]{Lu2019}
\bibinfo{author}{K.~Lu}, \bibinfo{author}{Y.~Jin}, \bibinfo{author}{Y.~Chen},
  \bibinfo{author}{Y.~Yang}, \bibinfo{author}{L.~Hou},
  \bibinfo{author}{Z.~Zhang}, \bibinfo{author}{Z.~Li}, \bibinfo{author}{C.~Fu},
\newblock \bibinfo{title}{{Review for order reduction based on proper
  orthogonal decomposition and outlooks of applications in mechanical
  systems}},
\newblock \bibinfo{journal}{Mechanical Systems and Signal Processing}
  \bibinfo{volume}{123} (\bibinfo{year}{2019}) \bibinfo{pages}{264--297}.
\bibitem[{Noor and Peterst(1980)}]{Noor1980}
\bibinfo{author}{A.~K. Noor}, \bibinfo{author}{J.~M. Peterst},
\newblock \bibinfo{title}{{Reduced Basis Technique for Nonlinear Analysis of
  Structures}},
\newblock \bibinfo{journal}{AIAA Journal} \bibinfo{volume}{18}
  (\bibinfo{year}{1980}) \bibinfo{pages}{455--462}.
\bibitem[{Idelsohn and Cardona(1985)}]{Idelsohn1985}
\bibinfo{author}{S.~R. Idelsohn}, \bibinfo{author}{A.~Cardona},
\newblock \bibinfo{title}{{A reduction method for nonlinear structural dynamic
  analysis}},
\newblock \bibinfo{journal}{Computer Methods in Applied Mechanics and
  Engineering} \bibinfo{volume}{49} (\bibinfo{year}{1985})
  \bibinfo{pages}{253--279}.
\bibitem[{Sombroek et~al.(2018)Sombroek, Tiso, Renson, and
  Kerschen}]{Sombroek2018}
\bibinfo{author}{C.~S. Sombroek}, \bibinfo{author}{P.~Tiso},
  \bibinfo{author}{L.~Renson}, \bibinfo{author}{G.~Kerschen},
\newblock \bibinfo{title}{{Numerical computation of nonlinear normal modes in a
  modal derivative subspace}},
\newblock \bibinfo{journal}{Computers and Structures} \bibinfo{volume}{195}
  (\bibinfo{year}{2018}) \bibinfo{pages}{34--46}.
\bibitem[{Jain et~al.(2017)Jain, Tiso, Rutzmoser, and Rixen}]{Jain2017}
\bibinfo{author}{S.~Jain}, \bibinfo{author}{P.~Tiso}, \bibinfo{author}{J.~B.
  Rutzmoser}, \bibinfo{author}{D.~J. Rixen},
\newblock \bibinfo{title}{{A quadratic manifold for model order reduction of
  nonlinear structural dynamics}},
\newblock \bibinfo{journal}{Computers and Structures} \bibinfo{volume}{188}
  (\bibinfo{year}{2017}) \bibinfo{pages}{80--94}.
\bibitem[{Benner et~al.(2015)Benner, Gugercin, and Willcox}]{Benner2015}
\bibinfo{author}{P.~Benner}, \bibinfo{author}{S.~Gugercin},
  \bibinfo{author}{K.~Willcox},
\newblock \bibinfo{title}{{A Survey of Projection-Based Model Reduction Methods
  for Parametric Dynamical Systems}},
\newblock \bibinfo{journal}{SIAM Review} \bibinfo{volume}{57}
  (\bibinfo{year}{2015}) \bibinfo{pages}{483--531}.
\bibitem[{{Baur Peter Benner Bernard Haasdonk Christian Himpe Immanuel Maier
  Mario Ohlberger} and {Dynamik Komplexer}(2017)}]{BaurBenner2017}
\bibinfo{author}{U.~{Baur Peter Benner Bernard Haasdonk Christian Himpe
  Immanuel Maier Mario Ohlberger}}, \bibinfo{author}{F.~{Dynamik Komplexer}},
  \bibinfo{title}{{Comparison of methods for parametric model order reduction
  of instationary problems}}, \bibinfo{year}{2017}. \URLprefix
  \url{http://www.mpi-magdeburg.mpg.de/preprints/}.
\bibitem[{Xiao et~al.(2015)Xiao, Fang, Buchan, Pain, Navon, and
  Muggeridge}]{Xiao2015}
\bibinfo{author}{D.~Xiao}, \bibinfo{author}{F.~Fang}, \bibinfo{author}{A.~G.
  Buchan}, \bibinfo{author}{C.~C. Pain}, \bibinfo{author}{I.~M. Navon},
  \bibinfo{author}{A.~Muggeridge},
\newblock \bibinfo{title}{{Non-intrusive reduced order modelling of the
  Navier-Stokes equations}},
\newblock \bibinfo{journal}{Computer Methods in Applied Mechanics and
  Engineering} \bibinfo{volume}{293} (\bibinfo{year}{2015})
  \bibinfo{pages}{522--541}.
\bibitem[{Xiao et~al.(2017)Xiao, Fang, Pain, and Navon}]{Xiao2017}
\bibinfo{author}{D.~Xiao}, \bibinfo{author}{F.~Fang}, \bibinfo{author}{C.~C.
  Pain}, \bibinfo{author}{I.~M. Navon},
\newblock \bibinfo{title}{{A parameterized non-intrusive reduced order model
  and error analysis for general time-dependent nonlinear partial differential
  equations and its applications}},
\newblock \bibinfo{journal}{Computer Methods in Applied Mechanics and
  Engineering} \bibinfo{volume}{317} (\bibinfo{year}{2017})
  \bibinfo{pages}{868--889}.
\bibitem[{Xiao(2019)}]{Xiao2019}
\bibinfo{author}{D.~Xiao},
\newblock \bibinfo{title}{{Error estimation of the parametric non-intrusive
  reduced order model using machine learning}},
\newblock \bibinfo{journal}{Computer Methods in Applied Mechanics and
  Engineering} \bibinfo{volume}{355} (\bibinfo{year}{2019})
  \bibinfo{pages}{513--534}.
\bibitem[{Zimmermann(2019)}]{Zimmermann2019}
\bibinfo{author}{R.~Zimmermann},
\newblock \bibinfo{title}{{Manifold interpolation and model reduction}}
  (\bibinfo{year}{2019}) \bibinfo{pages}{1--36}.
\bibitem[{Barrault et~al.(2004)Barrault, Maday, Nguyen, and
  Patera}]{Barrault2004}
\bibinfo{author}{M.~Barrault}, \bibinfo{author}{Y.~Maday},
  \bibinfo{author}{N.~C. Nguyen}, \bibinfo{author}{A.~T. Patera},
\newblock \bibinfo{title}{{An ‘empirical interpolation' method: application
  to efficient reduced-basis discretization of partial differential
  equations}},
\newblock \bibinfo{journal}{Comptes Rendus Mathematique} \bibinfo{volume}{339}
  (\bibinfo{year}{2004}) \bibinfo{pages}{667--672}.
\bibitem[{Chaturantabut and Sorensen(2010)}]{Chaturantabut2010}
\bibinfo{author}{S.~Chaturantabut}, \bibinfo{author}{D.~C. Sorensen},
\newblock \bibinfo{title}{{Nonlinear Model Reduction via Discrete Empirical
  Interpolation}},
\newblock \bibinfo{journal}{SIAM Journal on Scientific Computing}
  \bibinfo{volume}{32} (\bibinfo{year}{2010}) \bibinfo{pages}{2737--2764}.
\bibitem[{Phalippou et~al.(2020)Phalippou, Bouabdallah, Breitkopf, Villon, and
  Zarroug}]{Phalippou2020}
\bibinfo{author}{P.~Phalippou}, \bibinfo{author}{S.~Bouabdallah},
  \bibinfo{author}{P.~Breitkopf}, \bibinfo{author}{P.~Villon},
  \bibinfo{author}{M.~Zarroug},
\newblock \bibinfo{title}{{‘On-the-fly' snapshots selection for Proper
  Orthogonal Decomposition with application to nonlinear dynamics}},
\newblock \bibinfo{journal}{Computer Methods in Applied Mechanics and
  Engineering} \bibinfo{volume}{367} (\bibinfo{year}{2020})
  \bibinfo{pages}{113120}.
\bibitem[{Cho et~al.(2020)Cho, Shin, Kim, and Cho}]{Cho2020}
\bibinfo{author}{H.~Cho}, \bibinfo{author}{S.~J. Shin},
  \bibinfo{author}{H.~Kim}, \bibinfo{author}{M.~Cho},
\newblock \bibinfo{title}{{Enhanced model-order reduction approach via online
  adaptation for parametrized nonlinear structural problems}},
\newblock \bibinfo{journal}{Computational Mechanics} \bibinfo{volume}{65}
  (\bibinfo{year}{2020}) \bibinfo{pages}{331--353}.
\bibitem[{Kast et~al.(2020)Kast, Guo, and Hesthaven}]{Kast2020}
\bibinfo{author}{M.~Kast}, \bibinfo{author}{M.~Guo}, \bibinfo{author}{J.~S.
  Hesthaven},
\newblock \bibinfo{title}{{A non-intrusive multifidelity method for the reduced
  order modeling of nonlinear problems}},
\newblock \bibinfo{journal}{Computer Methods in Applied Mechanics and
  Engineering} \bibinfo{volume}{364} (\bibinfo{year}{2020})
  \bibinfo{pages}{112947}.
\bibitem[{Hesthaven and Ubbiali(2018)}]{Hesthaven2018}
\bibinfo{author}{J.~S. Hesthaven}, \bibinfo{author}{S.~Ubbiali},
\newblock \bibinfo{title}{{Non-intrusive reduced order modeling of nonlinear
  problems using neural networks}},
\newblock \bibinfo{journal}{Journal of Computational Physics}
  \bibinfo{volume}{363} (\bibinfo{year}{2018}) \bibinfo{pages}{55--78}.
\bibitem[{Maulik et~al.(2020)Maulik, Lusch, and Balaprakash}]{Maulik2020}
\bibinfo{author}{R.~Maulik}, \bibinfo{author}{B.~Lusch},
  \bibinfo{author}{P.~Balaprakash},
\newblock \bibinfo{title}{{Reduced-order modeling of advection-dominated
  systems with recurrent neural networks and convolutional autoencoders}},
\newblock \bibinfo{journal}{arXiv}  (\bibinfo{year}{2020}).
\bibitem[{Astolfi(2008)}]{Astolfi2008}
\bibinfo{author}{A.~Astolfi},
\newblock \bibinfo{title}{{Model reduction by moment matching for nonlinear
  systems}},
\newblock in: \bibinfo{booktitle}{2008 47th IEEE Conference on Decision and
  Control}, volume \bibinfo{volume}{2015-Febru}, \bibinfo{publisher}{IEEE},
  \bibinfo{year}{2008}, pp. \bibinfo{pages}{4873--4878}. \URLprefix
  \url{http://ieeexplore.ieee.org/document/7039956/
  http://ieeexplore.ieee.org/document/4738791/}.
  \DOIprefix\doi{10.1109/CDC.2008.4738791}.
\bibitem[{Astolfi(2010)}]{Astolfi2010}
\bibinfo{author}{A.~Astolfi},
\newblock \bibinfo{title}{{Model Reduction by Moment Matching for Linear and
  Nonlinear Systems}},
\newblock \bibinfo{journal}{IEEE Transactions on Automatic Control}
  \bibinfo{volume}{55} (\bibinfo{year}{2010}) \bibinfo{pages}{2321--2336}.
\bibitem[{Ionescu and Astolfi(2016)}]{Ionescu2016}
\bibinfo{author}{T.~C. Ionescu}, \bibinfo{author}{A.~Astolfi},
\newblock \bibinfo{title}{{Nonlinear moment matching-based model order
  reduction}},
\newblock \bibinfo{journal}{IEEE Transactions on Automatic Control}
  \bibinfo{volume}{61} (\bibinfo{year}{2016}) \bibinfo{pages}{2837--2847}.
\bibitem[{Rafiq and Bazaz(2020)}]{Rafiq2020}
\bibinfo{author}{D.~Rafiq}, \bibinfo{author}{M.~A. Bazaz},
\newblock \bibinfo{title}{{A framework for parametric reduction in large-scale
  nonlinear dynamical systems}},
\newblock \bibinfo{journal}{Nonlinear Dynamics} \bibinfo{volume}{102}
  (\bibinfo{year}{2020}) \bibinfo{pages}{1897--1908}.
\bibitem[{Acar and Shkel(2008)}]{Acar2008}
\bibinfo{author}{C.~Acar}, \bibinfo{author}{A.~Shkel}, \bibinfo{title}{{MEMS
  Vibratory Gyroscopes}}, \bibinfo{publisher}{Springer}, \bibinfo{year}{2008}.
\bibitem[{Izadi et~al.(2018)Izadi, Braghin, Giannini, Milani, Resta, Brunetto,
  Falorni, Gattere, Guerinoni, and Valzasina}]{Izadi2018}
\bibinfo{author}{M.~Izadi}, \bibinfo{author}{F.~Braghin},
  \bibinfo{author}{D.~Giannini}, \bibinfo{author}{D.~Milani},
  \bibinfo{author}{F.~Resta}, \bibinfo{author}{M.~F. Brunetto},
  \bibinfo{author}{L.~G. Falorni}, \bibinfo{author}{G.~Gattere},
  \bibinfo{author}{L.~Guerinoni}, \bibinfo{author}{C.~Valzasina},
\newblock \bibinfo{title}{{A comprehensive model of beams' anisoelasticity in
  MEMS gyroscopes, with focus on the effect of axial non-vertical etching}},
\newblock \bibinfo{journal}{5th IEEE International Symposium on Inertial
  Sensors and Systems, INERTIAL 2018 - Proceedings}  (\bibinfo{year}{2018})
  \bibinfo{pages}{1--4}.
\bibitem[{Wang et~al.(2018{\natexlab{a}})Wang, Phlipot, Perez, and
  Mignolet}]{Wang2018a}
\bibinfo{author}{X.~Q. Wang}, \bibinfo{author}{G.~P. Phlipot},
  \bibinfo{author}{R.~A. Perez}, \bibinfo{author}{M.~P. Mignolet},
\newblock \bibinfo{title}{{Locally enhanced reduced order modeling for the
  nonlinear geometric response of structures with defects}},
\newblock \bibinfo{journal}{International Journal of Non-Linear Mechanics}
  \bibinfo{volume}{101} (\bibinfo{year}{2018}{\natexlab{a}})
  \bibinfo{pages}{1--7}.
\bibitem[{Wang et~al.(2018{\natexlab{b}})Wang, O'Hara, Mignolet, and
  Hollkamp}]{Wang2018b}
\bibinfo{author}{X.~Q. Wang}, \bibinfo{author}{P.~J. O'Hara},
  \bibinfo{author}{M.~P. Mignolet}, \bibinfo{author}{J.~J. Hollkamp},
\newblock \bibinfo{title}{{Reduced Order Modeling with Local Enrichment for the
  Nonlinear Geometric Response of a Cracked Panel}},
\newblock \bibinfo{journal}{AIAA Journal} \bibinfo{volume}{57}
  (\bibinfo{year}{2018}{\natexlab{b}}) \bibinfo{pages}{421--436}.
\bibitem[{Budiansky(1967)}]{Budiansky1967}
\bibinfo{author}{B.~Budiansky},
\newblock \bibinfo{title}{{Dynamic Buckling of Elastic Structures: Criteria and
  Estimates}},
\newblock in: \bibinfo{booktitle}{Proceedings of an International Conference
  Held at Northwestern University, Evanston, Illinois},
  \bibinfo{publisher}{Pergamon Press Ltd}, \bibinfo{year}{1967}. \URLprefix
  \url{http://linkinghub.elsevier.com/retrieve/pii/B9781483198217500107}.
  \DOIprefix\doi{10.1016/B978-1-4831-9821-7.50010-7}.
\bibitem[{Wang et~al.(2013)Wang, Cen, and Li}]{Wang2013}
\bibinfo{author}{X.~Wang}, \bibinfo{author}{S.~Cen}, \bibinfo{author}{C.~Li},
\newblock \bibinfo{title}{{Generalized neumann expansion and its application in
  stochastic finite element methods}},
\newblock \bibinfo{journal}{Mathematical Problems in Engineering}
  \bibinfo{volume}{2013} (\bibinfo{year}{2013}).
\bibitem[{Hay et~al.(2010)Hay, Borggaard, Akhtar, and Pelletier}]{Hay2010}
\bibinfo{author}{A.~Hay}, \bibinfo{author}{J.~Borggaard},
  \bibinfo{author}{I.~Akhtar}, \bibinfo{author}{D.~Pelletier},
\newblock \bibinfo{title}{{Reduced-order models for parameter dependent
  geometries based on shape sensitivity analysis}},
\newblock \bibinfo{journal}{Journal of Computational Physics}
  \bibinfo{volume}{229} (\bibinfo{year}{2010}) \bibinfo{pages}{1327--1352}.
\bibitem[{Tiso(2011)}]{Tiso2011}
\bibinfo{author}{P.~Tiso},
\newblock \bibinfo{title}{{Optimal second order reduction basis selection for
  nonlinear transient analysis}},
\newblock in: \bibinfo{booktitle}{Proceedings of the 29th IMAC A Conference on
  Structural Dynamics 2011}, volume~\bibinfo{volume}{3}, \bibinfo{year}{2011},
  pp. \bibinfo{pages}{27--39}. \URLprefix
  \url{http://link.springer.com/10.1007/978-1-4419-9299-4
  http://link.springer.com/10.1007/978-1-4419-9299-4{\_}3}.
  \DOIprefix\doi{10.1007/978-1-4419-9299-4_3}.
\bibitem[{Krack and Gross(2019)}]{Krack2019}
\bibinfo{author}{M.~Krack}, \bibinfo{author}{J.~Gross},
  \bibinfo{title}{{Harmonic Balance for Nonlinear Vibration Problems}},
  \bibinfo{year}{2019}. \DOIprefix\doi{10.1007/978-3-030-14023-6}.
\bibitem[{Jutho et~al.(2019)Jutho, getzdan, Lyon, Protter, S, Leo, Garrison,
  Otto, Saba, Iouchtchenko, Privett, and Morley}]{jutho_2019_3245497}
\bibinfo{author}{Jutho}, \bibinfo{author}{getzdan}, \bibinfo{author}{S.~Lyon},
  \bibinfo{author}{M.~Protter}, \bibinfo{author}{M.~P. S},
  \bibinfo{author}{Leo}, \bibinfo{author}{J.~Garrison},
  \bibinfo{author}{F.~Otto}, \bibinfo{author}{E.~Saba},
  \bibinfo{author}{D.~Iouchtchenko}, \bibinfo{author}{A.~Privett},
  \bibinfo{author}{A.~Morley}, \bibinfo{title}{Jutho/tensoroperations.jl:
  v1.1.0}, \bibinfo{year}{2019}. \URLprefix
  \url{https://doi.org/10.5281/zenodo.3245497}.
  \DOIprefix\doi{10.5281/zenodo.3245497}.
\bibitem[{Woiwode et~al.(2020)Woiwode, Balaji, Kappauf, Tubita, Guillot,
  Vergez, Cochelin, Grolet, and Krack}]{woiwode:hal-02424746}
\bibinfo{author}{L.~Woiwode}, \bibinfo{author}{N.~N. Balaji},
  \bibinfo{author}{J.~Kappauf}, \bibinfo{author}{F.~Tubita},
  \bibinfo{author}{L.~Guillot}, \bibinfo{author}{C.~Vergez},
  \bibinfo{author}{B.~Cochelin}, \bibinfo{author}{A.~Grolet},
  \bibinfo{author}{M.~Krack},
\newblock \bibinfo{title}{{Comparison of two algorithms for Harmonic Balance
  and path continuation}},
\newblock \bibinfo{journal}{{Mechanical Systems and Signal Processing}}
  \bibinfo{volume}{136} (\bibinfo{year}{2020}) \bibinfo{pages}{106503}.

\end{thebibliography}

\end{document}